# Socially minded intelligence: How individuals, groups, and artificial intelligence can make each other smarter (or not)

Bingley, W. J.[1,2], Haslam, S. A.[2], Wiles, J.[3]

1. Faculty of Education, The University of Melbourne
2. School of Psychology, The University of Queensland
3. School of Electrical Engineering and Computer Science, The University of Queensland

[*]Corresponding author: Dr William J. Bingley: william.bingley@unimelb.edu.au, Faculty of Education, The University of Melbourne, Carlton South, Victoria, 3053.

## Author Note

The published version of this paper can be found in Personality and Social Psychology Review: https://doi.org/10.1177/10888683261421666. This research was jointly funded by the Australian Research Council Centre of Excellence for the Dynamics of Language (CE140100041), a University of Queensland strategic grant on Human-Centred AI, and a UQ Cyber grant on Cyber Security for Digital Identities. William J. Bingley is the recipient of an Office of National Intelligence National Intelligence Postdoctoral Grant (project number NIPG202413) funded by the Australian Government.

**Academic Abstract**

A core part of human intelligence is the ability to work flexibly with others to achieve goals. The incorporation of artificial agents into human spaces is making increasing demands on artificial intelligence (AI) to demonstrate and facilitate this ability. However, this kind of flexibility is not well understood because existing approaches to intelligence typically construe this either as an individual-difference trait or as a property of groups. We argue that by focusing either on individual or collective intelligence without considering their dynamic interaction, existing conceptualizations of intelligence limit the potential of people and AI systems. To address this impasse, we propose a new kind of intelligence—*socially minded intelligence*—that can be applied to both individuals and collectives. We outline how socially minded intelligence might be measured and cultivated within people, how it might be modelled in AI agents, and how it might be applied to other intelligent systems.

**Public Abstract**

In psychology, ‘intelligence’ is generally understood to be something that either individuals or groups have. However, the extent to which people can make each other more intelligent by working collectively—and the extent to which groups are smarter for having individuals who can think for themselves—is underexplored. Artificial intelligence (AI) research has a similar problem, meaning that artificial agents lack the ability to engage in this kind of intelligence, both with each other and with people. To address this gap in the literature, we outline a new kind of intelligence for psychology and AI—*socially minded intelligence*—which can be applied to individuals, groups, and artificial agents. We discuss how socially minded intelligence might be measured, improved, modelled in AI agents, and applied to other intelligent systems such as teams consisting of people and AI agents.

**Socially minded intelligence: How individuals, groups, and artificial intelligence can make each other smarter (or not)**

Imagine the following scenario: You are stranded with a small group of strangers on a deserted island after your intercontinental flight was forced to make an emergency crash landing. Furthermore, imagine that your primary goal—what you want more than anything else—is to get home alive. How might each of the following features of this scenario help you achieve this goal? (A) you have a high intelligence quotient (IQ); (B) you and the other survivors of the crash immediately form a cohesive, yet intellectually diverse group; or (C) you and the other survivors are able to act as individuals, subgroups, or as a single unit depending on the demands of the situation.

Individual perspectives on intelligence would say that Feature A (IQ) will be important, but that Features B and C are irrelevant to understanding your intelligence because they are contingent on other people rather than being individual traits or capabilities. Collective perspectives on intelligence would say that Feature B (group cohesion combined with cognitive diversity) is important for the group's ability to achieve its goals—but these may not overlap with *your* goals as an individual. In the present paper we focus on Feature C, which relies on an individual capability (being able to flexibly act as an individual or a group member), but also on other individuals and the social context. Regarding your personal goal of getting home alive, Feature C allows you either to work with the group if they share your goal, or to split off on your own if the group decides they want to stay on the island forever. Moreover, Feature C also contributes something unique beyond Feature B to the group's collective intelligence, as it allows the group to do such things as split into different foraging and exploration subgroups, or for subgroups to challenge the group's consensus in decision-making. In short, Feature C—the ability of agents to switch between acting as individuals or as group members depending on the demands of the situation—is a basis for both you and your group to have *socially minded intelligence*.

We propose that socially minded intelligence is a key feature of intelligence more broadly (i.e., both for individuals and collectives), but that this concept has been neglected in social science and artificial intelligence (AI) research because researchers have tended to focus *either* on individual intelligence *or* on collective intelligence without considering their dynamic interaction. In the present paper we address this lacuna by first identifying an appropriate definition of intelligence that can encompass both individual and collective intelligence in different kinds of agents. We then review literature on the intelligence of humans (individuals and groups) and artificial systems (single- and multi-agent) before outlining and defining the interactionist concept of socially minded intelligence and deriving relevant metrics. We conclude by exploring applications of this concept.

**Positionality Statement**

The authors of this paper are all White and hold academic positions at a university in a wealthy, White-majority country (Australia). Their social environment has shaped both their views on intelligence as a concept, and indeed has shaped how their own 'intelligence' has been understood and enabled. Regarding the first point, the first two authors take a social identity approach to psychology (Tajfel & Turner, 1979; Turner et al., 1987), which emphasizes the importance of considering relations between groups and group members in understanding psychological phenomena. This metatheoretical approach was developed explicitly to challenge more individualistic conceptualizations of psychology (see Reicher et al., 2012). Moreover, the third author has worked extensively on topics where the importance of social contexts for individuals is highly relevant, such as social robotics, natural language, and human-centred computing. Accordingly, the specific research backgrounds of the authors have informed their conceptualization of intelligence as a social process in the present paper.

Furthermore, throughout their lives the authors have benefitted from working in a world in which their intelligence has been judged in relation to norms, standards and metrics that reflect the norms and values of White, wealthy Westerners—particularly as these are framed by a cultural conceptualization of intelligence that sees this as an individual trait existing independently of the social environment. The perceived value of this kind of intelligence within the cultural context they live in has provided the authors with a range of benefits, including career opportunities and a legitimated voice on issues akin to those discussed in the present paper. Beyond these privileges, and in line with the broader definition of intelligence used in this paper, their intelligence in practice has been amplified by formal and informal assistance from other people, including senior collaborators and students. Their ability to achieve their goals has been facilitated further by groups, organizations and institutions—lab groups, universities, and funding agencies. These social resources have amplified their intelligence in practice (i.e., goal achievement) beyond their individual abilities. The intention of this paper is not just to acknowledge that existing intelligence metrics are often biased (see Gould, 1981; Richardson, 2002; Sternberg & Grigorenko, 2004), but also to capture how intelligence is *enabled* by the efforts of other people and groups and to make this social component of intelligence visible and distinct.

**Defining Intelligence**

There are many ways to define intelligence, with this concept having been applied to a diverse range of systems (Sternberg, 2020b). In the present paper we seek to understand the intelligence not only of humans, but also of AI systems: 'intelligent machines, especially intelligent computer programs" (McCarthy, 2004, p. 2). We focus on both human and artificial intelligence for several reasons. First, these kinds of intelligence are increasingly integrated through the practice of embedding AI in tools and systems (Afroogh et al., 2024; Gillespie et al., 2023), the growing implementation of AI agents within human environments (Casheekar et al., 2024; Gao et al., 2024)), and the development of human-AI teaming (O'Neill et al., 2022; Schmutz et al., 2024). Second, there is a deep two-way relationship between psychology and AI, with a long history of the former shaping the latter (Barto et al., 1981; Hebb, 1949; Hinton & Salakhutdinov, 2006; Krizhevsky et al., 2012; Laird, 2012; Legg & Hutter, 2007; McCulloch & Pitts, 1943; Oettinger, 1952; Turing, 1948, 1950; Widrow & Hoff, 1960) and vice versa (Anderson, 1990; Casey & Moran, 1989; Goel & Davies, 2020; Neisser, 1967; Newell & Simon, 1972; Solso, 1988). Third, although these fields have different aims—with psychology generally seeking to explain and facilitate existing intelligence and AI often focusing on creating new kinds of intelligence—both share the same broad conceptualization of intelligence as something that either individuals or groups have independently. Accordingly, both fields share the same blind spot (focusing either on individual or collective intelligence without considering their dynamic interaction)—and therefore the same *opportunity* that might be addressed simultaneously through conceptual reframing.

### *Agent-Based Intelligence*

With this scope in mind, we take a perspective on intelligence that is *agent-based*—where an agent is "anything that can be viewed as perceiving its environment through sensors and acting upon that environment through actuators" (Russell & Norvig, 2021, p. 51). Agent-based approaches have been used successfully to create both single-agent (Singh et al., 2022) and multi-agent (Gronauer & Diepold, 2022) artificial systems. They have also been applied to understand the behavior of both individual people (Scalco et al., 2019) and human groups (Goldstone & Janssen, 2005). Agent-based intelligence is defined by Legg and Hutter (2007, p. 402) as "an agent's ability to achieve goals in a wide range of environments". This definition of intelligence is intended to represent "the essence of intelligence in its most general form" (Legg & Hutter, 2007, p. 402) and can therefore be used to understand the intelligence of both people and AI systems. Furthermore, we define 'environment', specifically performance or task environment, as the problem(s) faced by an agent in achieving its goals (following Russell & Norvig, 2021, p. 40).

In the present paper we base our definition of intelligence on that of Legg and Hutter but adjust the framing of this in two key ways. First, we broaden the term 'agent' to include multi-agent systems, not just individual agents. Specifically, we propose that because multi-agent systems are capable of having higher-level goals (Braubach et al., 2005; Dorri et al., 2018) and have the potential to act in a coordinated manner in their environment (Cao et al., 2013), Legg and Hutter's definition can be applied to a *group* of agents to the extent that they (a) have shared higher-level goals and (b) act in their environment as a collective. This conceptualization of groups as agents is similar to that proposed by a number of other researchers (e.g., Bratman, 2013; Malone, 2018).

Second, for our present purpose we assume that relevant measures of the intelligence of agents and multi-agent systems will capture their performance in relation to specific goals (whether these are desired states of the world, or more open-ended drives to maximize or minimize a particular state). In other words, we exclude the process of *choosing* high-level goals. However, we include selection of new subgoals as a valid way of achieving existing higher-level goals, so long as the latter still exist to measure the current state of goal attainment. This restriction somewhat limits our definition of intelligence as it applies to humans, as goal setting and goal adaptation are considered to be part of human intelligence (Sternberg, 2020a). We make this simplification in order to be able to apply a similar standard to measuring the intelligence of both humans and agent-based AI systems (which have specified high-level goals in the form of desired states, utility functions, or performance standards; Russell & Norvig, 2021). Indeed, without this exclusion of choosing high-level goals from intelligence, no existing AI system would be considered intelligent (Fjelland, 2020; Goertzel, 2014). An implication of this approach is that the measurement of intelligence requires us to specify the goals that it corresponds to (Malone, 2018).

Taking all the above considerations into account, for the purposes of the present paper, our working definition of intelligence extends that of Legg and Hutter (2007) to suggest that *intelligence is the ability of an agent or multi-agent system to achieve specific goals in a wide range of environments*. Following this definition, a person or group is considered intelligent to the extent that they can achieve specific goals across different environments. This conceptualization aligns with the way that intelligence is understood across the various literatures we are discussing. For example, commonly-used individual-difference measures such as IQ scores are intended to predict goal attainment in a variety of domains (Kim, 2008; Kriegbaum et al., 2018; Strenze, 2007). Similarly, collective intelligence metrics are used to predict group goal attainment across different kinds of tasks (Riedl et al., 2021). Regarding artificial intelligence, an artificial agent or multi-agent system can be said to be intelligent to the extent that it can achieve specified goals (e.g., with reference to performance benchmarks) in a wide variety of environments. Accordingly, our chosen definition can be used to discuss intelligence in all of the different kinds of systems in which we are interested.

## Existing Approaches to Human and Artificial Intelligence

### *Approaches to Understanding Human Intelligence*

Researchers in psychology and AI have tended to focus either on either individual or collective intelligence, but not on the interaction between these levels of analysis. For example, regarding psychology, there are many theories that conceptualize intelligence in terms of differences between individuals. Some of these theories, such as the Cattell-Horn-Carroll theory (CHC; McGrew, 2009) focus on a single factor of intelligence ('*g*' in the CHC theory) within a hierarchy of lower-level factors. Others theories focus on specific kinds of intelligence, such as practical intelligence (Sternberg, 2000), emotional intelligence (Salovey & Mayer, 1990), social intelligence (Kihlstrom & Cantor, 2020), or cultural intelligence (Earley & Ang, 2003); or on combinations of intelligences (Gardner, 2011). A common feature of these approaches is that they treat intelligence as an individual-difference trait or capability—that is, as something that an individual person 'has an amount of' and that remains fairly stable across contexts. This perspective is epitomized by IQ tests, which ostensibly identify an individual's level of *g*, primarily for the purposes of comparing individuals to each other (Byington & Felps, 2010; Richardson, 2002).

In contrast to these individualistic accounts of intelligence, other researchers have focused on collective intelligence. This is generally conceptualized as a group-level variable and has been defined as "groups of individuals acting collectively in ways that seem intelligent" (Malone & Bernstein, 2015, p. 3). For example, Woolley and colleagues (2010) have identified a general factor called '*c*', which is analogous to *g* in the CHC theory of individual human intelligence. Several variables that impact on collective intelligence have been identified, including the abilities of group members (Adair et al., 2013; Woolley et al., 2010), the specific configuration of group members' traits (Aggarwal & Woolley, 2013; Williams Woolley et al., 2007), collective attention (Aggarwal & Woolley, 2013; Woolley, 2009a, 2009b), group structure (Girotra et al., 2010; Larson & Jr, 2009; Malone, 2018; Momennejad, 2021), and intragroup processes (Faraj & Xiao, 2006; Malone et al., 2010; Okhuysen & Bechky, 2009). However, while collective intelligence research points to ways in which a group might be more or less intelligent, it does not specify how acting collectively impacts on the intelligence of group members *themselves* (in terms of their own individual goals). At the same time, as things stand, it is not clear how the group-level factors that contribute to collective intelligence, such as group structures and intergroup processes, are mediated through individual-level psychological variables such as trust (Costa et al., 2001; De Jong et al., 2016) and social identity (Ellemers et al., 2004; S. A. Haslam, 2004b).

### ***Limitations of Existing Approaches to Human Intelligence***

**Limitations of Individualistic Approaches.** Many high-level constructs within human intelligence research (such as *g* and *c*), as well as more specific factors (such as processing speed and emotional intelligence), predict variance in performance outcomes (Luo et al., 2006; O'Boyle Jr. et al., 2011; Schmidt & Hunter, 2004; Woolley et al., 2010). For example, *c* was found by Woolley and colleagues (2010) to account for 43% of performance on a variety of group tasks. Similarly, basic cognitive processes such as processing speed have been found to be positively correlated with scholastic performance (Luo et al., 2006). However, regardless of the construct or chosen outcome, there is substantial variance in performance left unexplained by existing conceptualizations of human intelligence (Fox & Spector, 2000; Luo et al., 2003;

Woolley et al., 2010). We propose that additional variance might be accounted for by the ability of a person to solve problems by thinking and acting together with other people. Specifically, we suggest that being able to work with others (and having others to work with) will predict variance in an individual person's intelligence beyond that based on their individual-difference traits and capabilities. Moreover, we propose that the ability of people to work with others *in a flexible, context-sensitive way* will predict variance in collective intelligence when accounting for existing variables that predict this outcome.

Existing individual tests of human intelligence inherently reflect the influence of the social environment (Sternberg & Grigorenko, 2004). These tests measure intelligence within existing human social structures, such as organizations, schools, and cultures. However, they either ignore the input of these social structures (as in IQ tests; Richardson, 2002), or use them to fine-tune the testing context (as in the case of successful intelligence as conceptualized by Sternberg, 1999; or in social intelligence as conceptualized by Kihlstrom and Cantor, 2020). In either case, these tests fail to account for the contribution of the social environment *itself* to a person's intelligence. For example, regarding the first approach, asking two people from different cultures or socio-demographic backgrounds to complete an IQ test and then comparing their scores fails to account for the variance in their performance that arises from the fact that IQ tests have been designed to predict outcomes within a specific (i.e., Western, white collar) social context (Gould, 1981; Richardson, 2002), and hence the test is likely to undervalue the intelligence of people from other backgrounds (Greenfield, 1997). Practical intelligence methods address this problem by testing more context-specific outcomes (e.g., assessing domain-relevant tacit knowledge; Sternberg, 2020a). However, if a person tends to solve problems with others (regardless of their socio-cultural background), requiring them to perform an intelligence test without those others will also undervalue their intelligence—defined as their ability to achieve their goals in a wide range of environments. In contrast, another person who tends to solve problems by themselves may have their intelligence *overestimated* since the range of environments evaluated does not include those in which working with others would be advantageous. We propose that the social context should be accounted for in theories and tests of intelligence, and indeed that its influence should be measured as part of intelligence itself.

**Limitations of Approaches to Collective Intelligence.** The current lack of integration between individual and collective human intelligence limits our understanding of the intelligence of groups and group members in two key ways.

The first is that current conceptualizations of collective human intelligence cannot be used to understand how intelligent a given person is in the practical contexts of everyday life. In particular, while collective intelligence research has identified features of individual group members that can help a group achieve its goals (e.g., task-relevant abilities; Malone & Woolley, 2020), this perspective provides no insight into how working with a group might improve the intelligence of a group member in terms of their own individual goals. For example, factors that contribute to collective intelligence may not predict whether a researcher's personal goal to write highly-cited first-author papers will be better achieved by working with one lab group versus another—as this depends on more than just how effective those lab groups are as collectives. In particular, if one lab group is cohesive but distributes authorship evenly amongst group members, while another is fragmented but allows the researcher to claim first-authorship for more papers, then the researcher may be better able to achieve their individual goal in the 'less effective' group. Similarly, an elite sportsperson playing a team sport may be better able to achieve their personal goal of being acknowledged as the best player in the world on a worse team where they are able to shine individually, rather than on a better team where they are required to play a reduced role. In other words, we agree with Malone and Woolley's (2020, pp. 793–794) claim that "studying collective intelligence provides a link between cognitive psychology and high-level social, organizational, and economic processes" but observe that this link has tended to be one-way in intelligence research—that is, it focuses on the ways in which individual cognitive psychology impacts on collective processes. The inverse link from social processes to individual psychology is neither imagined nor explored in intelligence research—although it has been explored in other domains (e.g., group-based trust, Cruwys et al., 2021; the impacts of social rituals on cognition, emotion, and behavior, Hobson et al., 2018; social norm change as a function of group communication about prototypes, Hogg & Reid, 2006).

The second problem that arises from focusing on the group level of analysis concerns the lack of clarity around how intragroup dynamics shape collective intelligence (Janssens et al., 2022). In particular, while independent yet integrated decision-making has been highlighted as important for collective intelligence outcomes (Caruso & Williams, 2008; Surowiecki, 2004; Woolley, 2009b), and different kinds of intragroup decision-making structure have been identified (Malone, 2018), the *dynamics* of these processes in terms of changes over time have not been clearly articulated. For example, the intermediate level of analysis between individuals and groups consists of subgroups that can (and often will) have their own subgoals that may contribute to or hamper the achievement of the overall goals of the group (Cronin et al., 2011). Accordingly, the extent to which group members see themselves and act as subgroup rather than group members in different contexts needs to be analyzed to determine the ability of the group to function as a whole (Bezrukova et al., 2009; Cooper et al., 2014; Dovidio et al., 2009). In some contexts group members may benefit the group best by acting more as differentiated individuals rather than as undifferentiated group members (Adelman & Dasgupta, 2019; Packer, 2008). For example, an individual group member going against the group's consensus (when this consensus is incorrect) may help to avoid problems such as groupthink; and having group members who retain some degree of individuality may allow them to contribute differing viewpoints, potentially increasing the group's intelligence (Nemeth et al., 2001). Finally, because different decision-making structures are optimal in different circumstances (Malone, 2018), being able to change between these structures may help groups be more intelligent (Lei et al., 2016; Siegel Christian et al., 2014). Existing accounts of collective intelligence do not have an effective way of conceptualizing and predicting these kinds of dynamics.

### ***Approaches to Artificial Intelligence***

While there are many subfields and approaches within the field of AI, we propose that there is, broadly speaking, a gap between research into single- and multi-agent AI that mirrors the gap between individual and collective intelligence research in psychology. For example, many tests of AI system intelligence focus on performance against individual human benchmarks (Bringsjord & Schimanski, 2003; Mahoney, 1999). These approaches are examples of 'psychological AI', which focuses on designing systems that think like (individual) humans (Goel & Davies, 2020). More broadly, though, even researchers in AI who do not measure system performance by comparison to individual humans have generally conceptualized the intelligence of artificial systems as an individual-difference variable—that is, as something a given system has an amount of by itself, which is relatively stable and can be compared against other systems across contexts (e.g., Chauhan et al., 2024; Desnoyer & Wettergreen, 2010; Hernández-Orallo, 2017). Moreover, many AI systems have been inspired by ideas from individualistic paradigms within psychology such as associative learning and cognitive psychology (Anderson, 1990; Goel & Davies, 2020; Hinton & Salakhutdinov, 2006; Laird, 2012; R. S. Sutton & Barto, 1981), propagating an implicitly individual model of human intelligence in AI. Accordingly, there is a perspective within AI that implies that the best way to improve a system's intelligence is to change its internal processes or components in some way (e.g., by improving processing speed, cognitive architecture, neural network size and depth, reinforcement learning, algorithm optimization, and training datasets; Bostrom, 2014; Cong & Zhou, 2023; Russell & Norvig, 2021). In other words, this perspective implies that the way to make AI systems more intelligent (as with people) is to make them smarter *as individuals*.

A system of cooperating artificial agents (a 'multi-agent system' or MAS) is comparable to a human group comprised of cooperating group members (Chmait et al., 2016), and the ability of a MAS to achieve its goals can be thought of as its collective intelligence. One approach to designing MAS, inspired by game theory, is to coordinate agents by utilizing utility calculations. For example, this can involve aligning agents' individual utility calculations to achieve equilibria (Stankovic et al., 2012; Tang & Yi, 2023), optimizing group-level utility (Caillou et al., 2010; Cui et al., 2012; Rădulescu et al., 2019; Gronauer & Diepold, 2022), or combining different types of equilibria-seeking behaviors (Semsar-Kazerooni & Khorasani, 2009; Verbeeck et al., 2002).

Another approach to designing MAS, 'swarm AI', is to construct systems that take inspiration from animals and other non-human biological systems (particularly ants and bees; Chakraborty & Kar, 2017) to produce emergent collective behavior (Askay et al., 2019; Bonabeau et al., 2020; Gray et al., 2018; Innocente & Grasso, 2019; Meng et al., 2016; Neshat et al., 2014). Game theoretic and swarm approaches can be combined in a single MAS (Hahn et al., 2020). Agents in MAS are generally designed as components, with the aim of making the overall system more intelligent (Russell & Norvig, 2021). This aim is reflected in the assessment of collective artificial intelligence, which tends to focus on the goals of the system as a whole (e.g., Sami, 2020; Sathiyaraj & Bharathi, 2020). Accordingly, MAS agents are often designed to be *less* individually intelligent than they would be if they were independent agents. As Altshuler (2023, p. 4) puts it, "while designing intelligent swarm systems we must assume (and often even aspire for) having … available individual agents that are myopic, mute, senile and rather stupid". Indeed, the kind of individualized thinking that typifies much human intelligence in groups is seen by some MAS researchers as a threat to coordination (L. Rosenberg & Willcox, 2020). In contrast to the uniform, simple agents in the majority of approaches to MAS intelligence, some studies have looked to investigate individuality, subgroups, and intragroup dynamics within collective systems (Aydin, 2012; Dorri et al., 2018; List et al., 2008). However, such approaches have focused primarily on the impact of these dynamics on *collective* (not individual) outcomes.

In summary, AI researchers have generally built and tested AI systems either as individual agents or as MAS. In the following section we discuss some of the limitations of this exclusive focus on either the individual or the collective level of analysis.

### *Limitations of Existing Approaches to Artificial Intelligence*

**Limitations of Individualistic Agents.** Just as the individualistic approach to intelligence poses problems for psychology, it also causes issues for AI. For example, one of the most apparent differences between biological and artificial intelligence is that the former tends to be general and the latter tends to be narrow (Neftci & Averbeck, 2019). This means that while AI systems are often very powerful, they are at the same time fragile and rely on humans to construct and maintain their performance environment, in ways that limit their intelligence in the sense that we have defined it in this paper (Brachman & Levesque, 2022). This is most apparent in contexts of use in which AI systems act with a high degree of autonomy in unconstrained settings involving interactions between agents (e.g., self-driving cars and other social robots). Such contexts can present a potentially infinite number of novel challenges outside the scope of any finite set of rules, where a human is not available to simplify the environment.

Progress in such fields, and towards solving the broader problem of 'common sense' in AI, has been slow compared with other areas where humans use narrow AI systems as tools (Brachman & Levesque, 2022; Shanahan et al., 2020; Zhu et al., 2020). An individualistic approach to improving general AI (e.g., Yampolskiy, 2012, p. 99; Bostrom, 2014; Sandberg & Bostrom, 2008) assumes that the way to increase the intelligence of an agent is to improve something about that agent itself (e.g., its architecture, model size, reward function, data set). This approach is epitomized in the design of artificial agents, in that other agents are typically considered as part of the problem to solve, rather than a way to solve the problem—unless the agent is explicitly designed to be part of a multi-agent system (e.g., Gronauer & Diepold, 2022). In contrast, most humans have the capacity to form part of a multi-agent team when this is appropriate or required to solve a given problem, and indeed this allows them to do so across varied environments (Malone, 2018; E. A. Smith, 2010; von Hippel, 2018). In other words, a human who can work with others around them does not necessarily need to change anything about themselves to be better able to achieve their goals. More generally, many other naturally occurring intelligent systems, including animals, bacteria, and even cancers, rely on higher-level cooperation to increase the ability of agents to reach their goals (Capp et al., 2023; Cheney, 2011; Lee et al., 2022; Sachs et al., 2004; Whiten et al., 2021). Along these lines, we propose that individual artificial agents might perform better on benchmarks of goal attainment across different environments if they were able to work with other agents to achieve their own goals when appropriate and/or necessary.

**Limitations of Multi-Agent Systems.** Existing conceptualizations of intelligence also limit the potential of MAS in two important ways. The first results from the fact that treating collective intelligence as a coordination problem is computationally demanding (Russell & Norvig, 2021). In particular, approaches which aim to coordinate the 'selfish' motivations of individual agents by establishing equilibria become intractable as the number of agents expands (Han & Hu, 2020; Papadimitriou, 2015; Wooldridge, 2009). This scalability problem also affects multi-agent reinforcement learning (Gronauer & Diepold, 2022; Mguni et al., 2018). In contrast, human cooperation can be scaled up to allow cooperation on a scale of hundreds, thousands, or millions of agents (Apicella & Silk, 2019). The second limitation is that, as with models of collective human intelligence, multi-agent system designs tend to be static rather than dynamic, in the sense that the relationships between agents and the groups they belong to do not change in response to circumstances. For example, MAS typically have fixed organizational structures, such as hierarchical ("agents have tree-like relations") or holonic ("agents are organized in multiple groups which are known as holons based on particular features, e.g., heterogeneity"), each of which is better suited to particular kinds of tasks (Dorri et al., 2018, p. 28585). By comparison, human cooperation is fluid and context-sensitive (Rand et al., 2011). This means that while a group may have a particular structure at a given time, this is subject to change depending on the context and the task at hand (Hagemann et al., 2012; Stachowski et al., 2009). These two features of human cooperation—scalability and flexibility—help people to solve problems in changing environments in ways that artificial MAS currently cannot.

In sum, failing to account for the dynamic interaction between individuals and groups leads to both overestimation and underestimation of the intelligence of people and human groups. The lack of such abilities in artificial systems also reduces the design space and hence potential intelligence of artificial agents and multi-agent systems. These, we suggest, are problems that a model of socially minded intelligence can help us address and resolve.

**Socially Minded Intelligence**

Thus far, our review of the literature on intelligence in individuals and groups (human and artificial) has revealed factors that contribute to individual intelligence (i.e., the ability of individuals to achieve goals) and collective intelligence (the ability of collectives to achieve goals) but it has also identified limitations that result from failing to account for dynamic interactions between these levels of analysis. In this section we propose a new kind of intelligence to capture these interactions. We call this *socially minded intelligence*. This kind of intelligence only exists in contexts containing more than one agent.

For a target agent (individual or group) we define socially minded intelligence as *the extent to which the agent (for an individual) or component agents (for a group) can flexibly perceive, think, and act alone or collectively towards the goals of the target agent.* In contexts with more than one agent, individual agents with socially minded abilities have the potential (but not the requirement) to form an intersubjective 'social mind' with others (Turner & Oakes, 1997). Such a social mind is a dynamic assembly of individual agents, each with their own individual goals, who in the present moment define themselves, and act, as group members and work towards shared goals. It is important to note that being able to perceive, think and act with others *requires others to be present*—which means that other agents are inherently part of a target agent's socially minded intelligence.

At the individual level, agents that can flexibly participate in a social mind with others have a theoretically increased potential to achieve their goals across different environments relative to agents that can only operate either as individuals or group members. We make this argument on the basis that a socially minded agent can achieve its goals in environments where this kind of flexibility is *required*, whereas an agent lacking this intelligence cannot. This therefore increases the number of potential problems that can be addressed. For example, a person caught in a violent political protest with the goal of survival who is able to join either the protestors or the authorities (or act as an individual) has more potential avenues for achieving their goal than a person who can only see themselves as a protestor, or only as an individual. Similarly, at the collective level, groups made up of such agents have an increased potential to achieve their goals relative to groups made up of agents that work only as individuals or only as group members. For example, the protestors in the same hypothetical situation have an increased number of future states in which they can achieve their collective goal of social change if they can temporarily act as individuals rather than always having to act as protestors. We argue that a similarly expanded set of goal-achievement states exists for socially minded artificial agents in environments requiring joint action for success populated by other agents that can be flexibly friendly, hostile, or neutral.

Indeed, there is evidence from the psychological literature to suggest that socially minded intelligence may have benefits in a range of domains required for intelligence (Malone, 2018), including learning, remembering, sensing, creating possibilities, and deciding on action. For example, individuals can learn more effectively with the help of others (Bentley, 2019; Vygotsky, 1978); other people can act as aids to memory (Lanzi et al., 2017; Momennejad, 2021; J. Sutton et al., 2010); creativity is unlocked in particular social contexts (S. A. Haslam et al., 2013); and decision-making can be improved through interaction with others (Gigerenzer, 2020; Mercier & Sperber, 2018). Collective intelligence might also be improved through the ability of group members to act flexibly as individuals or in subgroups. For example, having group members who can operate with some degree of independence can improve the ability of a group to generate ideas (Bissola & Imperatori, 2011; M. Rosenberg et al., 2022) and to decide on an effective course of action (Juni & Eckstein, 2015; Kameda et al., 2022; Surowiecki, 2004). Moreover, pro-social dissent within groups (whether by individuals or subgroups) can improve collective intelligence by challenging and changing harmful social norms (Packer, 2008) and improving creativity (De Dreu & West, 2001).

In light of the potential for collectives to contribute to individual intelligence and individuals to contribute to collective intelligence, we therefore propose that a new concept of socially minded intelligence might explain unique variance in the performance of individuals and groups. In this sense, we conceptualize socially minded intelligence as a phenomenon that *contributes* to, rather than constitutes, individual and collective intelligence. In other words, we assume that socially minded individuals and groups have a degree of intelligence (a capacity to achieve goals across different environments) that is independent of socially minded intelligence, and that socially minded intelligence therefore impacts on—rather than constitutes—this capacity. This contribution may be positive (e.g., when an individual forms a social mind with others that moves them towards their individual goal) or negative (e.g., when group members act as two subgroups with opposing goals to that of the group as a whole).

***Socially Minded Intelligence for Individual Agents***

Given our definition of intelligence as the ability to achieve goals in different environments, socially minded intelligence is formulated differently when viewed from an individual versus a collective perspective, since the goals of these entities may differ. The socially minded intelligence of an individual agent is defined in relation to the individual's goals as follows:

> *An agent's individual socially minded intelligence is the extent to which it can flexibly perceive, think, and act with other agents towards its own individual goals.*

An agent with a high level of socially minded intelligence must (a) be able to flexibly perceive, think, and act with other agents using its own social abilities; (b) have other agents in its social environment to perceive, think, and act with; and (c) be able to act towards its own individual goals together with those other agents. In contrast, an agent that lacks socially minded intelligence might (a) be unable to flexibly perceive, think, and act with other agents due to a lack of social capability; (b) not have other agents in their social environment to perceive, think, and act with; or (c) be unable to act towards its own individual goals with those other agents. In other words, the socially minded intelligence of a given agent is a function of the agent itself, other agents, their relationship, and the alignment between agents' individual and shared goals.

### ***Socially Minded Intelligence for Collectives***

We define a group as *a social self-category that is superordinate to the individual* (following Turner et al., 1987; Wiles et al., 2022). Further, 'social self-category' is defined as a social category that is subjectively self-relevant to at least one agent (through explicit self-representation or encoded implicitly); and 'superordinate to the individual' means higher-order (i.e., potentially applying to at least two agents). Examples of such groups might be a work team comprised of individual team members, a nation made up of citizens, or an emergent 'minimal' group based a single shared characteristic such as shared fate or interests. A group may look like a single individual from the 'outside', but that individual may be subjectively pursuing the goals of a group even above their own individual goals. For example, despite its official disbandment in 1945, the Imperial Japanese Army existed until 1974 in the mind of Hiroo Onoda, a Japanese soldier who did not surrender after World War 2 and continued to fight until 1974, including several years completely alone. The superordinate group in this case was the Imperial Japanese Army, of which he was the only member. Nevertheless, this group shaped his behavior such that he attacked others on the basis of being enemy combatants (opponents of this group). This can be contrasted with his subsequent behavior as an individual, and as a representative of a different group (e.g., Japan as a nation in peacetime). Moreover, based on our definition of 'group', individual agents do not need to interact to constitute a group. They *do* need to share an identity or goals (to the extent that these constitute a superordinate social category) – but they do not need to *know* reflexively that they share these. For example, people can see themselves as fans of a sporting team (a superordinate group) without necessarily interacting or being aware of other fans.

The collective socially minded intelligence of a group is defined in relation to the group's goals as follows:

> *A group's collective socially minded intelligence is the extent to which individual agents in the group can flexibly perceive, think, and act as group members, subgroup members, or individuals towards the group's superordinate goals.*

Accordingly, a group with a high degree of collective socially minded intelligence must be comprised of agents that (a) can perceive, think, and act as individuals or group members; (b) can change between these states dynamically; and (c) in doing so can act towards the group or system's superordinate goals. A group that lacks socially minded intelligence might be comprised of agents that (a) can perceive, think, and act either as individual or group members, but not both; (b) can perceive, think, and act either as individuals or group members but are not able to move between these states in response to the situation; or (c) in flexibly acting as individuals or group members fail to act towards the system's overall goals or act in ways that work against these goals. Accordingly, the socially minded intelligence of a group is determined by the perceptual, cognitive, emotional, and behavioral flexibility of component agents; and the extent to which this flexibility leads to actions which align the agents with the goals of the group.

There is a fundamental asymmetry to our conceptualizations of individual and collective socially minded intelligence. Specifically, we treat both individual and collective socially minded intelligence as operating through the internal states of individual agents. This is because while an individual agent can be intelligent without a collective, a group cannot be intelligent (in the sense of achieving its goals) without its members. Indeed, a group is only an agent at all to the extent that it has members who can perceive, think, and act *together*. Moreover, the potential for a group to be an agent in this sense only exists when its members perceive it *as a group*, which is again mediated through individual states. Accordingly, we refer to the internal states of individual agents, rather than an internal state of the group itself, when discussing collective socially minded intelligence. In other words, we treat collective socially minded intelligence as an emergent property of group members and their relation to the group.

### ***Stable, Variable, and Transferable Components of Socially Minded Intelligence***

Inherent in the definitions of socially minded intelligence outlined above are two key elements: a relatively *stable component* (i.e., agents' capacity to be socially minded that comes from their own individual-difference abilities), and a *variable component* (i.e., the extent to which other socially minded agents are present, and their self-concept and goal overlap with the target). This variable component can also be called the target agent's 'socially minded context', and it reflects the extent to which other agents have shifted from being part of the problem to be solved (i.e., part of the performance environment) to being part of problem-solving (i.e., part of the target agent's ability to achieve its goals). These elements combine in a given context to produce socially minded intelligence, following the dynamic person **x** situation conceptualization of interactionist social psychology (e.g., Turner et al., 1987). Moreover, there is the potential for a third element that is important for intelligence as we have defined it in this paper: a *transferable component*. This component occurs when a socially minded context remains stable across performance environments, such as when a cohesive human group stays together to solve problems in changing situations. In other words, socially minded intelligence always contains a relatively stable component and a situationally determined component, with the latter potentially being transferrable. These components are outlined in reference to the definitions of socially minded intelligence in Table 1.

The stable component of socially minded intelligence is represented by the individual-difference ability to be socially minded—to be able to flexibly perceive, think, and act with others—present in the target agent (for individual socially minded intelligence) or in the individual agents who make up a group (for collective socially minded intelligence). As an individual-difference ability, this component exists even when only a single socially minded agent is present. However, in such a situation it is not predicted to impact on that agent's intelligence because it has nothing to interact with. Moreover, even when other socially minded agents are present it is not the case that this ability will necessarily combine with the social context to increase a person or group's intelligence, as other socially minded agents may have goals that contradict the target's, leading to an overall negative effect on their intelligence. Accordingly, this stable component does not represent socially minded intelligence *per se*.

Rather, its impact on intelligence in a given context depends on the variable component, and therefore it is not attempting to explain the same variance in intelligence as other conceptually stable intelligence constructs such as *g* and *c*. Indeed, an individual who always perceives, thinks, and acts with others is not necessarily more intelligent as a result, and may in some circumstances be less intelligent. In this sense, their intelligence is *more dependent* on others—such that we propose that more socially minded agents have a 'higher ceiling and lower floor' for their intelligence compared with less socially minded agents.

Because socially minded intelligence contains a variable component that interacts with the stable component to impact on intelligence, it might be argued that such a construct is too contextual to have predictive value across performance environments. Accordingly, it is important to restate that a socially minded context can be transferrable to the extent that it does not change with the performance environment. And, to the extent that it is transferable, the variable component is predicted to impact systematically on the target agent's intelligence across performance environments, thereby explaining unique variance in the target agent's ability to solve problems. However, the socially minded context does not *have* to be stable. Indeed, it is precisely because socially minded context can be transferrable—*but need not necessarily be*—that socially minded intelligence is theoretically context-sensitive in a way that other intelligences are not. In other words, a transferrable socially minded context is like a tool that can be picked up and used to solve different problems by socially minded agents, but can potentially be abandoned if it is useless or counterproductive. We explore how some individuals might be better or worse at selectively utilizing this tool (which is different from their ability to use it at all) in the section 'Factors Impacting on Socially Minded Intelligence'.

**Table 1**

*Stable, Variable, and Transferrable Components of Socially Minded Intelligence.*

| Concept | Stable component | Variable component | Transferrable component |
|---|---|---|---|
| Socially minded intelligence | The individual-difference ability of an agent (for a target individual) or abilities of component agents (for a target group) to flexibly perceive, think, and act alone or collectively | The target's *'socially minded context'*, i.e., the extent to which multiple socially minded agents are present, and their degree of self-concept and goal overlap with the target | The extent to which the target's socially minded context is stable across performance contexts |
| Individual socially minded intelligence ('ISMI' in Box 1) | The individual-difference ability of a target agent to flexibly perceive, think, and act with other agents ('SMA' in Box 1) | The target agent's *'socially minded context'*, i.e., the extent to which other socially minded agents are present, and their degree of self-concept and goal overlap with the target agent ('SR' in Box 1) | The extent to which the target agent's socially minded context is stable across performance contexts |
| Collective socially minded intelligence ('GSMI' in Box 2) | The individual-difference ability of agents in a target group to flexibly perceive, think, and act as group members, subgroup members, or individuals ('SMA' in Box 2) | The target group's *'socially minded context'*, i.e., the extent to which group members' self-concepts and goals overlap with the group ('SO' and 'GO' respectively in Box 2) | The extent to which the target group's socially minded context is stable across performance contexts |

Note. The overall concept of socially minded intelligence (row 1) is superordinate to individual socially minded intelligence (row 2) and collective socially minded intelligence (row 3). Individual and collective socially minded intelligence are defined in relation to a specific target agent or group.

### ***Examples of Socially Minded Intelligence***

So far, we have defined socially minded intelligence in abstract terms so that it can be applied to people, AI systems, individuals and groups. To make the concept more concrete and distinct from other intelligence constructs, we can consider hypothetical examples of what socially minded intelligence might look like. For example, an elite sporting or military team may have a great deal of collective intelligence in their normal sphere of operation—and indeed team members may also have high levels of individual intelligence. However, such a team may not have much individual or collective socially minded intelligence if group members are 'fused' (Swann Jr. et al., 2009) such that they permanently see themselves in terms of this collective identity (as this would reduce the self-definitional flexibility of group members). Accordingly, such a group may be high performing in its regular performance environment, but both individuals and the group itself may be susceptible to environmental change that requires individual action or action in terms of a different collective identity. For example, members of such a group and the group itself would be predicted to struggle in peacetime, or during ceasefires (for military units); or in the offseason, in other levels of competition (e.g., domestic vs international), and in retirement (for athletes). If the stable component of socially minded intelligence were greater for such a group, this would predict that the group would find such transitions easier. For example, an elite military team that has high socially minded intelligence would be predicted to be able to disband more effectively during ceasefires or peacetime because members can act as individuals or in terms of other social identities as needed. Moreover, such a group would also be able to psychologically reform as a group quickly if hostilities resume.

On the other hand, a group might have a great deal of socially minded intelligence without a high level of other individual and collective factors that predict intelligence. For example, a group of people with average IQ scores and without a formal structure or task-relevant abilities may nevertheless be able to adapt to a changing environment (for their individual and collective benefit) so long as they have a high level of socially minded intelligence. Here, the extent to which the variable component of socially minded intelligence is both positive and transferrable will determine whether an otherwise unremarkable group can flourish in changing environments. Needless to say, a group with low individual, collective, and socially minded intelligence is not predicted to flourish in any circumstance.

***The Dynamic Interactionist Metatheory of Socially Minded Intelligence***

Socially minded intelligence has a *dynamic interactionist methatheory* (Turner & Oakes, 1986, 1997), which considers that understanding people and groups requires considering how "the mind (e.g. thoughts, emotions, memory, perception, imagination) and mental functioning are qualitatively transformed… through social interaction and shared activities" (Reynolds et al., 2010, p. 466). This is contrasted against an individualistic or reductionist metatheory, "where explanations of such collective phenomena are reduced to asocial, non-contextualised, abstract causes (e.g. temperament, mood, biology, individual differences, limitations and biases of cognition)". The dynamic interactionist metatheory, as represented historically by the social psychology of Sherif (1936) Lewin (1952), Asch (1952), and researchers in the social identity tradition (Tajfel, 1979; Turner & Oakes, 1997), builds on principles of Gestält cognitive psychology and applies these ideas to cognition about people (Turner et al., 1987). Specifically, this kind of interactionism explains psychological phenomena as resulting from a dynamic interaction between a person and their situation, where neither the person nor their situation are fully independent and can be understood in isolation (Reynolds et al., 2010). Accordingly, socially minded intelligence can be considered dynamically interactionist in that it refers to the part of a person or group's intelligence that is formed in a given situation, depending on the interaction between qualities of individuals (the stable component) and the dynamic relations between individuals and groups (the variable/transferrable component). This interaction constitutes socially minded intelligence, not the components by themselves.

There are several other psychological concepts, traditionally understood using an individualistic metatheory, that can similarly be reconceptualized using a dynamic interactionist metatheory. For example, the 'standard theory' of power (Fiske & Berdahl, 2007; Keltner et al., 2003; Raven, 1993) is that it is the capacity to influence other people based on control over outcomes. However, it has been argued (Simon & Oakes, 2006; Turner, 1991, 2005) that this kind of power is better described as 'power over', with another kind of power ('power through', or 'social power') following not from potential coercion but from the recruitment of human agency via shared identity. A similar analysis has been applied to leadership, reconceptualizing this not as "a solo process but one that is grounded in relationships and connections between leaders and those they influence (i.e., followers)" (S. A. Haslam et al., 2024, p. 3). This social identity approach to leadership (S. A. Haslam et al., 2020) argues that leadership can be understood not only in terms of individual qualities of leaders or their formal position within hierarchies, but also in terms of the dynamic interaction between leaders and followers. This is not to say that an identity leader has no agency in their ability to lead, but rather that this ability is *contingent* on their social context. Similarly, although socially minded intelligence is contingent on the social context, individuals and groups may seek to maintain or increase their socially minded intelligence and transfer this across performance environments. Other psychological phenomena that have been reconceptualized using an interactionist metatheory in a similar way include personality (Reynolds et al., 2010; Turner & Onorato, 1999) and creativity

(S. A. Haslam et al., 2013). In considering how intelligence can be determined by the dynamic interaction between individuals and groups, the concept of socially minded intelligence follows this tradition.

## Socially Minded Intelligence in People

In this section we discuss how the concept of socially minded intelligence might be conceptualized, measured, and improved for people and the groups they belong to.

### *The Relationship Between Individual, Collective, and Socially Minded Intelligence*

As previously stated, socially minded intelligence is considered to contribute to, rather than constitute, the intelligence of individuals and collectives. In other words, socially minded intelligence is not individual or collective intelligence *per se*, but rather an interactive process that contributes to both of these as outcomes. In this sense, socially minded intelligence is similar to individual and collective *intelligences* (see Figure 1), whether these are unitary concepts such as *g* or *c* or more specific intelligences such as linguistic intelligence or logical-mathematical intelligence. The primary difference between these existing intelligences and socially minded intelligence is that the former are generally conceptualized as either individual- or collective-level variables, whereas socially minded intelligence is conceptualized as an interaction between individual and collective-level variables. Indeed, it is *because* it accounts for the dynamic interaction between individuals and groups that we propose that socially minded intelligence contributes unique variance to individual and collective intelligence beyond that explained by existing intelligences.

**Figure 1**

Individual, Collective, and Socially Minded Intelligence

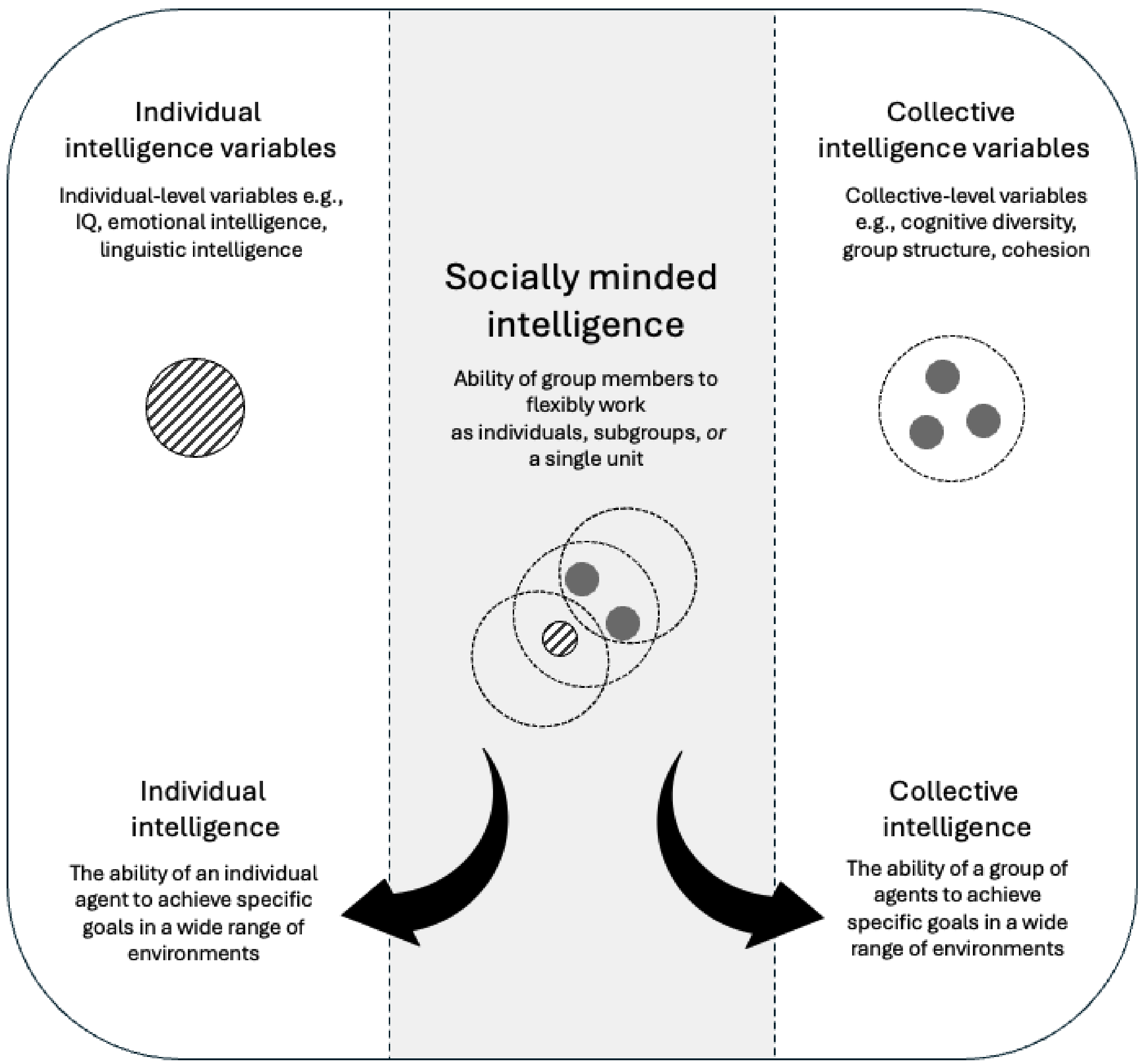

*Note.* This figure represents the unique contribution to individual and collective intelligence represented by the abilities of individuals, groups, and the interactionist concept of a ‘social mind’ formed by individuals who can act as group members or subgroup members when required (socially minded intelligence).

**Differences From Individual Intelligences.** Socially minded intelligence is distinct from the other kinds of individual-level intelligences that we reviewed earlier in this paper in that it is not a relatively stable individual-difference variable, but rather a function of individual-difference and contextual inputs. It shares similarities with social intelligence (Conzelmann et al., 2013; Kihlstrom & Cantor, 2020) to the extent that this ability allows an agent to connect with others, but social intelligence by itself is missing the variable/transferrable component of socially minded intelligence that allows the latter to vary contextually. Accordingly, a person could have a great deal of social intelligence but not have any socially *minded* intelligence because they lack others to perceive, think, and act with—or because acting with those others goes against their personal goals. Another similar individual-difference intelligence is Sternberg's (2003, p. 141) concept of successful intelligence—"the ability to achieve success in life in terms of one's personal standards, within one's sociocultural context". Socially minded intelligence differs from this in conceptualizing the social (although not cultural) context as contributing to intelligence through its dynamic interaction with the individual. Ultimately, because socially minded intelligence is partially contextually determined, it does not provide the same information as existing individual intelligences, such as a person's ability to achieve goals without working with others. In this sense, it is unfair to use a person's socially minded ability as an indication of their purely individual ability. Indeed, it is not intended to explain the part of a person's intelligence that comes from their abilities in isolation. Rather, it is intended to represent a person's ability to be intelligent towards their own goals *with* other people *in situ*.

More broadly, considering how socially minded intelligence might contribute to a person's individual intelligence enables a reevaluation of the tendency of people to conform or agree with others (e.g., Cialdini & Goldstein, 2004; Hoffman et al., 2001; Wood, 2000). For example, people tend to trust ingroup members more than outgroup members, even when this trust is based on arbitrary groups and even when it can lead to risk-taking behavior (Cruwys et al., 2021; Voci, 2006). In experiments to extract the influence of social environment from cognition, such tendencies appear to reduce individual intelligence (Clegg et al., 2017; N. Rhodes & Wood, 1992; Zuckerman et al., 2013). However, variables such as ingroup trust, cohesion, and the ability of group members to be influenced by leaders are vital for group dynamics, not least because they enable people to achieve goals collectively (Ellemers et al., 2004; van Vugt & Hart, 2004; Wilmot & Ones, 2022) and are thus an inherent component of their socially minded intelligence. Accordingly, from a socially minded intelligence perspective the individualized, static experimental context of psychology studies and intelligence tests is likely to overestimate the extent to which people who follow others are 'biased', rather than optimized for their social world. This prediction is consistent with research showing that heuristics (cognitive shortcuts) that lead to errors in experimental contexts can lead to successful performance in ecologically valid contexts (Gigerenzer, 2020; Oakes et al., 1994). Just as human rationality is less effective when deployed by individuals without the input of other people (Mercier & Sperber, 2018), so human intelligence more broadly appears impoverished when it is removed from its social context.

**Differences From Collective Intelligences.** Socially minded intelligence also differs from existing factors that contribute to collective intelligence because it is not (a) a fixed quality of group members or a configuration of traits (Adair et al., 2013; Williams Woolley et al., 2007; Woolley et al., 2010); (b) a fixed quality of the group itself (Girotra et al., 2010; Larson & Jr, 2009; Malone, 2018; Momennejad, 2021); or (c) a quality of group dynamics explained at the group level (e.g., Faraj & Xiao, 2006; Malone et al., 2010; Okhuysen & Bechky, 2009). Rather, it connects individual qualities (specifically, the ability to be socially minded) with group-level factors through consideration of the psychological relations between group members and the group in a given context. This means that it can explain how intragroup processes such as leadership and competition interact with the psychology of group members to shape collective intelligence *dynamically*. For example, a socially minded intelligence analysis predicts that a group where group members have a high (versus low) average level of socially minded ability has the potential for more trust, cohesion, and coordination—but also the potential for *less*, depending on the salience of individual or subgroup identities and the degree to which their goals overlap with that of the group. In other words, there is no such thing as an inherently intelligent group in a socially minded sense—it depends on the dynamic interaction between individual group members and their socially minded context.

Moreover, socially minded intelligence further differs from other collective intelligence factors in that it can also be used to explain *how collectives impact on the intelligence of individuals*. For example, an individual group member's potential to achieve their goals in different environments is predicted to be maximized in a context where they can form a social mind intersubjectively with others—if this social mind is directed towards their own individual goals. In other words, socially minded intelligence links individual psychology to group factors in a contextual, bidirectional way that other collective intelligence factors do not.

**Other Differences.** Both individual and collective socially minded intelligence also differ conceptually from the *motivation* or drive to work collectively. In particular, individuals may have a greater or lesser level of socially minded intelligence with the same level of motivation to work collectively, depending on the alignment of their goals with the goals of other group members. Similarly, a group in which individuals are highly motivated to work collectively may not have a great deal of socially minded intelligence if group members have low socially minded abilities.

Furthermore, because socially minded intelligence requires consideration of the target's goals from the target's perspective it cannot be understood as a straightforward *extension* of the concepts of either individual or collective intelligence. For example, individual socially minded intelligence might potentially be conceptualized as simply expanding individual intelligence to account for the contributions of others. However, the individual socially minded intelligence of group members cannot be used to calculate the socially minded intelligence *of the group itself*, because the former does not consider the group's goals in its calculation. Similarly, a group's socially minded intelligence might be considered an expansion of the concept of collective intelligence in the sense that it includes the aligned contributions of individual group members. However, collective socially minded intelligence does not tell us how much socially minded intelligence a given group member has. Indeed, a group of strongly psychologically invested group members might hypothetically achieve its own goals at the *expense* of its group members' individual goal achievement. This further justifies why we have defined individual and collective socially minded intelligence separately.

### *Measuring Socially Minded Intelligence*

Measuring a person's socially minded intelligence requires new methods that capture not only their abilities as an individual, but also their social environment. As a starting point, a target person's ability to perceive, think, and act with others (when such others are available) can be operationalized using measures of their ability to flexibly construe the self in individual or collective terms. This is because seeing the self in collective terms allows individuals to integrate information from others into their perception of the world (Bentley et al., 2017; Coppin et al., 2016; Cunningham & Van Bavel, 2009; S. A. Haslam et al., 1996; Shankar et al., 2013; Van Bavel & Cunningham, 2012; Xiao & Van Bavel, 2012), to selectively incorporate others' inputs into their own decision-making (Cohen, 2003; Esposo et al., 2013; Fielding et al., 2020; Graupensperger et al., 2021; Turner, 1991; van Knippenberg et al., 1994; Voci, 2006), and to work with others towards shared goals (Ellemers et al., 2004; S. A. Haslam et al., 2000; S. A. Haslam & Reicher, 2012; van Knippenberg, 2000; van Knippenberg & Ellemers, 2003). Critically, in order to have socially minded intelligence, collective self-construal must be *flexible*, so that individuals are not 'stuck' at a particular level of self-construal. One such construct that represents this flexibility is self-categorization ability (Skorich & Haslam, 2022). This and other similar abilities could be combined to form a composite measure of *socially minded ability, SMA*, representing the target person's individual-difference traits and skills that are important for socially minded intelligence. SMA is the quantitative operationalization of the stable component of socially minded intelligence.

The social environment of a target person provides two key variables that measure the extent to which other people will contribute towards the target's socially minded intelligence. The first of these reflects the extent to which each person present shares a sense of social identity with the target person (*shared social identity*, *SSI;* e.g., Khan et al., 2015). Because people are more trusting of (Cruwys et al., 2021; Voci, 2006), have better communication with (Greenaway et al., 2015, 2016), and are more well-intentioned towards those with whom they share a sense of social identity (Dovidio et al., 1997; Levine et al., 2005; Tajfel & Turner, 1979), shared social identity can be seen as representing the extent to which a person will perceive, think, and act with available others. The second relevant variable provided by the social environment is the extent to which the target person's personal goals align with those they might potentially work with in the current context. One way to quantify this variable is to measure the *goal alignment, GA,* between the target person and each other person in the context.

The *social resources, SR*, of the target person combines their shared social identity and goal alignment with each other person in the current context. This term represents how much the potential social power (Simon & Oakes, 2006) represented by shared social identity is enabled by goal alignment. SR is the quantitative operationalization of the target person's 'socially minded context'—the variable (and potentially transferrable) component that contributes to their socially minded intelligence in a given situation. A person's socially minded intelligence (the extent to which they can perceive, think, and act with other agents towards their own individual goals) is therefore a function of their *individual-difference socially minded ability* and *social resources, which in turn combines shared social identity* and *goal alignment* in that context. This *individual socially minded intelligence* (*ISMI*) is the quantitative operationalization of the interaction between the stable component of socially minded intelligence (i.e., SMA) and the variable/transferrable component (i.e., SR) in a given situation. Box 1 provides an example of how this measure might be calculated more specifically.

For groups, it is possible to calculate a similar measure of socially minded intelligence (the extent to which each person in the group can flexibly perceive, think, and act as a group member, subgroup member, or individual, and act accordingly towards the system's superordinate goals). This requires measuring (a) the individual-difference abilities of group members that are relevant to socially minded intelligence; (b) the degree to which the group is self-defining for group members (i.e., 'group identification'; Postmes et al., 2013); and (c) the extent to which the goals of the identity that is currently most salient to group members (e.g., individual, subgroup, superordinate group; Forehand et al., 2002) align with the group's goals. In other words, a *group's socially minded intelligence* (*GSMI*) is a function of *group members' individual-difference socially minded ability, SMA;* their *group identification, GI*; and their *salient identity goal alignment, SIGA*. As with ISMI, GSMI is the quantitative operationalization of the interaction between the stable (SMA) and variable/transferrable components (GI and SIGA) of socially minded intelligence in a particular situation. In this case, however, the target for which socially minded intelligence is calculated is a group rather than an individual agent. Box 2 provides an example of how GSMI might be calculated more specifically in a given context.

The definitions for ISMI and GSMI derived in Boxes 1 and 2 show a deep mathematical similarity between their formulations. The general concepts of aligned abilities, self-overlap, and goal overlap may also be useful for defining other ways to use the concepts of ISMI and GSMI (see Box 3).

***Box 1. Individual socially minded intelligence in a given context***

***Definition**. A person's ISMI is the extent to which they can flexibly perceive, think, and act with other people towards their own individual goals. (From p. 22)*

In what follows we derive a measure of individual socially minded intelligence *(ISMI)* from its natural language definition and desired computational properties.

A positive socially minded intelligence value for an individual represents the extent to which the target person will be perceiving, thinking, and acting with other agents *towards* their own individual goals by working with others in the context; while a negative socially minded intelligence value represents the extent to which the target person will be perceiving, thinking, and acting with other agents *away* from their own individual goals by working with others in the context. Note that ISMI operates differently to measures of individual intelligence such as IQ—which is a stable indicator of a person's intelligence—because it is always a function of a given social context.

To define the ISMI metric for a given context, properties P1 – P3 define the relevant variables, and P4 – P6 provide the criteria for how they are combined:

P1. *SMA and SR:* Each individual has a *socially minded ability (SMA)*, which is assumed to be relatively stable across contexts. This ability contributes to achieving a person's goals when they interact with others, who constitute their *social resources (SR)*. SMA ranges from 0 (totally unable to perceive, think, and act with available others) to 1 (fully capable of perceiving, thinking, and acting with available others). ISMI is the product of SMA and SR. If there are no other people around, regardless of how much individual intelligence the target person has, their SR would be 0 and hence their ISMI would be 0.

P2. *SSI:* In a given context, the social identities of two people can vary, and we define their degree of *shared social identity (SSI)* ranging from 0 (no shared social identity) to 1 (fully shared social identity). Note that SSI can change in different contexts, even between the same two people, since contexts can make different social identities more or less salient.

P3. *GA:* In a given context, the goals of two people can vary from fully aligned to fully opposed, which we define as their degree of *goal alignment (GA)*. It is defined to range from -1 (goals are fully opposed), through 0 (goals are unrelated), to 1 (goals are fully aligned).

P4. A person who has a high SSI with the target person in the context will increase the target person's ISMI if their GA is positive (their goals are aligned) and decrease the target person's ISMI if their GA is negative (their goals are opposed). Note that this means that a target person's ISMI can be negative in contexts where others that they identify with strongly have opposing goals.

P5. A person who has a low SSI with the target person in the context will have minimal or no effect on the target person's ISMI.

P6. To realize properties P4 – P5, the contribution of a person in the context (a *contributor*) to a target person's SR is calculated as the product, *SSI* x *GA*. Each additional person contributes independently to this measure.

We can now formally define the *ISMI* metric, for a target person, *p*, and a given context, *c*, as follows:

$$ISMI_{pc} = SMA_p * SR_{pc} \qquad \text{Equation (1)}$$

where $SMA_p$ is the target person's socially minded ability; and $SR_{pc}$ is the target person's social resources in the given context. $SMA_p$ is a non-negative real number, ranging from 0 – 1 (see P1 above). $SR_{pc}$ is calculated as the sum of shared social identity with all others, indexed by $q$, in the context, $c$, as follows:

$$SR_{pc} = \Sigma_q (SSI_{pqc} * GA_{pqc}) \qquad \text{Equation (2)}$$

where $SSI_{pqc}$ is shared social identity between the target individual $p$, and each other person, $q$ (where $q \neq p$) in context $c$. $SSI_{pqc}$ is defined as a non-negative value from 0 – 1 (see P2 above). $GA_{pqc}$ is the goal alignment between the target person, $p$, and each other person, $q$, in context, $c$, and is a real value from -1 to +1 (see P3 above).

**Worked Example 1. Calculating the contribution of one other person to *ISMI***

If a target person, $p$, has an $SMA$ of .7, and is in a context with one other person, $q$, who shares average levels of $SSI$ = .5 and moderately positive $GA$ =.5 with the target person, then the target person's $ISMI$ would be calculated using Equation (1) above as follows (results shown to 2 significant figures):

$$ISMI_{pc} = 0.7 * SR_{pc}$$

$$= 0.7 * \Sigma_q (SSI_{pqc} \text{ x } GA_{pqc})$$

$$= 0.7 * (0.5 * 0.5)$$

$$= 0.7 * 0.25$$

$$= 0.18$$

With the contribution of one other person, *ISMI* can vary from -1 to +1. Even the moderate levels of *SSI* and *GA* in this example contribute positively to the *ISMI* of the target person, albeit not by a large amount.

**Example Notes:** Each additional person in the context who has *SSI* > 0 adds to the social resources for the target person (see Appendix B, Examples B1, and B2 for worked examples for 2 – 5 other people). *ISMI* has additional interesting properties: Someone with a low *SMA* in a context with several people with average *SSI and GA* can have a greater *ISMI* than someone with a higher *SMA* but just one other person as their social resource (see Appendix B, Examples B3 and B4). *ISMI* can also be reduced by the presence of others, as shown in the following example:

**Worked Example 2. Calculating the contribution to *ISMI* from others with positive and negative *GA***

If a target person, *p*, has an *SMA* of .7, and is in a context with five other people who have varying levels of *SSI* (.5, .3, .4, .7, 0) and a range of positive and negative levels of *GA* (.5, .7, -.8, -.3, -.9), then the target person's SR and ISMI are affected by the presence of others with positive or negative GA to the extent that they have SSI with each other person, as shown in the following (results shown to 2 significant figures):

$$
\begin{aligned}
ISMI_{pc} &= 0.7 * SR_{pc} \\
&= 0.7 * \textstyle\sum_q (SSI_{pqc} \text{ x } GA_{pqc}) \\
&= 0.7 * ((0.5 * 0.5) + (0.3 * 0.7) + (0.4 * -0.8) + (0.7 * -0.3) + (0 * -0.9)) \\
&= 0.7 * (0.25 + 0.21 - 0.32 - 0.21 + 0)
\end{aligned}
$$

= -0.049

With the contribution of five other people, *ISMI* can vary from -5 to +5. This example shows how negative and positive social resources can counteract each other. Note that in this second Worked Example it is important to calculate the social resources for the target person from each person first before summing them (e.g. negative goal alignments are only detrimental if there is also a positive shared social identity with the target individual). In relatively homogeneous groups, the product of total SSI and average GA would be a reasonable approximation to a target person's social resources, but these approximations are less valid in heterogeneous groups.

***Box 2. Calculating a group's socially minded intelligence in a given context***

**Definition**. *GSMI is the extent to which group members can flexibly perceive, think, and act as group members, subgroup members, or individuals towards the group's superordinate goals. (From p. 23)*

The derivation of a measure of group socially minded intelligence *(GSMI)* follows a similar structure to Box 1. However, instead of evaluating others as social resources for a target individual (as in Box 1), the group measure here requires estimating the contribution of the *aligned abilities* of members towards the group as a whole. A positive socially minded intelligence value for a group thus represents the extent to which group members' flexibility in terms of self-definition contributes *towards* the group's goals in the present context; while a negative socially minded intelligence value represents the extent to which group members' flexibility in terms of self-definition works *against* the group's goals.

There is a subtle distinction in the concept of aligned abilities which is critical to the definition of GSMI: GSMI represents members' self-definitional flexibility, group identification, and salient identity goal alignment, *not* the group's goal-directed ability more generally (which would be collective intelligence).

To define the GSMI metric for a given context, properties P7 – P9 define the relevant variables, and P10 – P12 provide the criteria for how they are combined:

P7. *SMA and AA:* Each group member has a *socially minded ability (SMA)*, which is equivalent to their SMA as described in property P1. This ability contributes to achieving a group's goals to the extent that their identities and goals are aligned with the group, which we call their *aligned abilities (AA)*, and which can vary from -1 to +1. Note that AA from each contributor towards the group may differ from their AA with other members individually (this distinction is indicated by subscripts in the equations below). GSMI averages the aligned abilities of all the members in the group.

P8. *GI and SO:* In a given context, each group member will identify more or less with the group (i.e., the group will be more or less self-defining), and we define their degree of *group identification (GI)* ranging from 0 (no group identification) to 1 (group is totally self-defining). Note that GI shares similarities with SSI from Box 1 in that both are measures of *self-overlap (SO)* between a contributor and a target unit and can change in different contexts. They differ in the scale of their targets from individuals for ISMI and groups for GSMI and also in that self-overlap is intersubjective ('two-way') in the case of SSI, and unidirectional ('one-way', from the perspective of each group member) for GI.

P9. *SIGA and GO:* In a given context, the goals of the most salient identity for each group member (e.g., 'individual', 'subgroup', 'group') and the group can vary from fully aligned to fully opposed, which we define as their degree of *salient identity goal alignment (SIGA)*. This is defined to range from -1 (goals are fully opposed), through 0 (goals are unrelated), to 1 (goals are fully aligned). Note that SIGA shares similarities with GA from Box 1 in that both are measures of *goal-overlap (GO)* between a contributor and a target unit (individual vs group) and can change in different contexts. In the case of GA the overlap is between the goals of the target person and each contributing person, while for SIGA the overlap is between the goals of the group and the goals of the identity that is most salient to each contributing person. In both cases the overlap is two-way, because groups and individuals can both have goals. Like GA,

SIGA can change in different contexts as identities become more or less salient to each group member.

P10. Each person joining a group has the potential to increase, maintain or reduce the group's social resources: If everyone in a group has the same level of self-definitional flexibility (SMA), group identification, and goal alignment with the group, the number of people in the group does not impact on the group's average aligned abilities. As the group grows larger, the relative impact of each individual group member on the group's GSMI is reduced.

P11. If a person joins the group with below average aligned abilities (which could be due to low goal alignment *or* below average group identification), the group's GSMI is reduced.

P12. To realize properties P10-P11, the aligned abilities of a person in the context towards the group as the target unit is calculated as the product, *SMA* x *GI* x *SIGA*, which can be written more generally in terms of self- and goal overlaps as *SMA* x *SO* x *GO*.

The formal definition for GSMI can now be given as follows:

*GSMI*, is defined for a target group, *g*, in a given context, *c*, as the average across all group members of their aligned abilities, which is calculated as the product of three context-dependent factors: (1) each group member's individual-difference abilities relevant to the group's socially minded intelligence, $SMA_m$, (2) their group identification, $GI_m$, and (3) their salient identity goal alignment, $SIGA_m$:

$$GSMI_{gc} = (\Sigma_m (SMA_m * GI_m * SIGA_m))/N_m \quad \text{Equation (3)}$$

where $SMA_m$ is each group member's individual-difference socially minded abilities as defined in Equation 1, with the subscript $m$ indexed over all group members in the context, and $N_m$ is the number of group members.

**Worked Example 3. Calculating a group's *GSMI* in a context with 2 group members.**

In a context where a target group, *g*, has 2 group members present with varying *SMA*s (.7, .8), *GI*s (.3, .8) and alignment between each group member's current salient identity and the group's goals (.5, .3), the group's *GSMI* would be calculated using Equation (3) as follows (results shown to 2 significant figures):

$$GSMI_{gc} = (\Sigma_m (SMA_m * GI_m * SIGA_m))/N_m$$

$$= ((0.7 * 0.3 * 0.5) + (0.8 * 0.8 * 0.3))/2$$

$$= (0.11 + 0.19)/2$$

$$= 0.15$$

Regardless of the number of members in the group, *GSMI* can vary from -1 to +1. This example shows a relatively small positive value, reflecting the low alignments of the individuals with the group.

**Notes.** A group could in principle have a single member. In this case, the GSMI is the product of the single group member's individual SMA, their group identification, and their salient goal alignment with the group goals (see Appendix C Example C1). Example 3 extends naturally to averaging the contributions of additional group members (see Appendix C examples C2-C5).

**Worked Example 4. Calculating the contribution to *GSMI* from a group with positive and negative group goal alignment**

In a context where a target group, *g*, has 5 group members present with varying *SMA*s (.7, .8, .6, .9, .5), varying *GI*s, (.3, .8, .2, .7, .3) and both positive and negative alignment between each group member's current salient identity and the group's goals (.5, .3, -.8, .4, -.1), the group's *GSMI* would be calculated using Equation (3) above as follows (results shown to 2 significant figures):

$$GSMI_{gc} = (\Sigma_m (SMA_m * GI_m * SIGA_m))/N_m$$

$$= ((0.7 * 0.3 * 0.5) + (0.8 * 0.8 * 0.3) + (0.6 * 0.2 * -0.8) + (0.9 * 0.7 * 0.4) + (0.5 * 0.3 * -0.1))/5$$

$$= (0.11 + 0.19 – 0.096 + 0.25 – 0.015)/5$$

$$= 0.088$$

As in Example 3, GSMI can vary from -1 to +1. However, when more individuals with low alignment and/or a mixture of positive and negative goal alignment are part of the group, GSMI tends towards zero for the group as a whole in that context.

***Box 3. Towards a general formula for socially minded intelligence in a given context***

Based on the derivations in of ISMI and GSMI in Boxes 1 and 2, we can now use the concepts of *aligned abilities AA, self-overlap (SO), and goal overlap (GO)* to define similar forms of the equations for ISMI and GSMI.

The aligned abilities $AA_{xqc}$, to a target unit $x$ ($x = p$ for an individual; $x = g$ for a group), from a contributor, $q$, in a context $c$, can be expressed as:

$$AA_{xqc} = SMA_x * SO_{xqc} * GO_{xqc} \qquad \text{Equation (4)}$$

Combining Equations 1 – 2 from Box 1, the formulation for $ISMI_{pc}$ can be rewritten as the sum of the *aligned abilities* $(AA_{pqc})$ of the target person, $p$, from the contributors, $q$, in context $c$ as follows:

$$\begin{aligned} ISMI_{pc} &= SMA_p * \Sigma_q \, (SSI_{pqc} * GA_{pqc}) \\ &= \Sigma_q \, (SMA_p * SSI_{pqc} * GA_{pqc}) \\ &= \Sigma_q \, (SMA_p * SO_{pqc} * GO_{pqc}) \\ &= \Sigma_q \, (AA_{pqc}) \qquad \text{Equation (5)} \end{aligned}$$

Equation (5) has a parallel derivation in the formulation for the $GSMI_{gc}$ as the sum of the *aligned abilities* $(AA_{gqc})$ of the contributors, $q$, to the group $g$, in context, $c$:

$$\begin{aligned} GSMI_{gc} &= (\Sigma_q \, (SMA_q * GI_{gqc} * SIGA_{gqc})) / N_m \\ &= (\Sigma_q \, (SMA_q * SO_{gqc} * GO_{gqc})) / N_m \\ &= \Sigma_q \, (AA_{gqc}) / N_m \qquad \text{Equation (6)} \end{aligned}$$

The similarities between the forms of Equations (5) and (6) shows that the measurement of socially minded intelligence is not tied to a single level of analysis (individual or group) but can be applied across the different scales of target units. A similar formula could be derived for the intermediate sub-units of a group.

The equations in Boxes 1, 2, and 3 require that socially minded ability, self-overlap (group identification and shared social identity), and goal overlap are quantified in some way. While measures for some of these constructs, such as group identification (Leach et al., 2008; Postmes et al., 2013) and shared social identity (Khan et al., 2015) have been developed, others such as socially minded ability (which we have argued may be roughly represented by self-categorization ability; Skorich & Haslam, 2022) and identification of goals and their alignment (Seijts & Latham, 2000; Y. Zhang & Chiu, 2012) require further conceptualization and development.

### ***Factors Impacting on Socially Minded Intelligence***

We have previously discussed how socially minded intelligence differs from individual and collective intelligence, broadly speaking, in that it impacts on both of these as outcomes. Moreover, because socially minded intelligence is interactionist (i.e., it is constituted by the dynamic interaction between individuals and social contexts), it is also likely that individual and collective intelligence factors will have an impact on it in turn. We refer here to factors beyond those in the definitions outlined above (and their corresponding measurement equations), drawing primarily from research in psychology given that socially minded artificial agents do not yet exist. For each factor, we outline how these impacts might operate through the terms in the equations for calculating socially minded intelligence (see Boxes 1, 2, and 3 for explanations of these terms).

**Personality Traits.** One set of individual factors that may impact on socially minded intelligence are personality traits that systematically predispose someone towards working with others. For example, agreeableness ("individual differences in the motivation to maintain positive relations with others"; Graziano & M. Tobin, 2017, p. 105) and extraversion ("individual differences in the tendencies to experience and exhibit positive affect, assertive behavior, decisive thinking, and desires for social attention"; Wilt & Revelle, 2017, p. 57). These kinds of 'pro-social' factors can be considered as impacting on an individual's socially minded intelligence through systematically increasing self-overlap across situations, leading to a 'higher ceiling but lower floor' for their intelligence (higher and lower possible values of individual socially minded intelligence depending on goal overlap). While we do not predict that this will necessarily lead to higher individual socially minded intelligence, an inclination to work as a group member rather than an individual may predict higher individual socially minded intelligence on average, given the cooperative nature of much of human behavior (Boyd & Richerson, 2009). Indeed, both agreeableness (Wilmot & Ones, 2022) and extraversion (Wilmot et al., 2019) have been associated with positive outcomes for individuals in a range of domains.

**Individual-Difference Intelligences.** Social (Conzelmann et al., 2013; Kihlstrom & Cantor, 2020) and emotional (Kotsou et al., 2019; O'Boyle Jr. et al., 2011; Salovey & Mayer, 1990) intelligence may also shape socially minded intelligence to the extent that these improve an individual's ability to know when being socially minded will increase their own individual intelligence. This may be good for individuals but bad for collectives because of the interdependence of these levels of socially minded intelligence as alluded to in in Box 3. Specifically, an individual group member who optimizes their own self-overlap based on their level of goal overlap with others (i.e., higher shared social identity when goal alignment is high, lower shared social identity when goal alignment is low, no shared social identity when goal alignment is negative) will maximize their possible individual socially minded intelligence, but at the same time negatively impact on the group's socially minded intelligence when the group's goals differ from the individual's (i.e., when there is higher group identification when salient identity goal alignment is high, and lower group identification when salient identity goal alignment is low). In other words, making self-overlap *contingent* on goal overlap gives more power to individuals and less to groups—which may ultimately undermine not only collective intelligence, but also individual intelligence (if other agents stop being socially minded with the 'selfish' individual as a result). A promising individual-difference ability for resolving this tension is *wisdom*, which has been described as the ability to resolve conflicts 'among different intrapersonal, interpersonal, and extrapersonal (i.e., group-centric) interests in people's lives' (Grossman, 2017, p. 236). Wisdom may enable individuals to flexibly use socially minded intelligence in a way that accounts for both individual and collective responsibilities.

**Group Dynamics.** Another set of factors that may impact on socially minded intelligence relates to group dynamics. For example, the extent to which a particular social identity is made self-relevant to all group members in a given situation should increase self-overlap with other group members, as well as salient identity goal alignment. This may happen due to changes in the environment (e.g., in response to a shared threat; Brewer, 2000), but it could also be an *agentic* process. Indeed, increasing a sense of shared identity in their followers is something that identity leaders (S. A. Haslam et al., 2020) actively try to do. This is because, from an identity leadership perspective, self-overlap with other group members quantifies the degree to which a leader can be influential in the group (to the extent that they also embody how the group sees itself and are seen to be advancing its cause in the world; see Steffens et al., 2014). Accordingly, we predict that (effective) identity leadership will increase self-overlap within the group as well as salient identity goal alignment—thereby increasing the group socially minded intelligence of the group and either increasing or decreasing the individual socially minded intelligence of group members depending on their personal goal overlap with the group. *Intergroup* dynamics may also shape socially minded intelligence. For example, conflict between groups should increase ingroup identification within those groups, thereby shaping individual and group socially minded intelligence (Greenaway & Cruwys, 2019; Mead & Maner, 2012).

**Multiple Identities.** Another factor that may shape socially minded intelligence relates to the fact that people typically belong to multiple groups (Sønderlund et al., 2017), which may have more or less compatible goals. In the present paper, following self-categorization theory, we assume that these identities are *functionally antagonistic* such that when one becomes salient others become less so (Turner et al., 1987). Accordingly, this will then shape the shared social identity term for individuals and self-overlap for groups (as well as salient identity goal alignment when a new identity becomes the most salient). For example, in Worked Example 1, if a different identity that is not shared with the other person in the situation is made more salient for the target person, their shared social identity with that other person will decrease in inverse proportion to the salience of the new identity. Similarly, in Worked Example 3, if other identities become more salient to group members, this will reduce the salience of the overall group identity (reducing group members' group identification); and to the extent that these other identities become the *most* salient, this will also shape their salient identity goal alignment (depending on the goals associated with those identities). A further complication, which we do not capture in our current equations, is that self-overlap and goal overlap may not be completely independent, such that when an individual identifies with a group their *personal* goals may also change. This is a potential area for future empirical research and theoretical expansion.

**Information and Reasoning.** Finally, if aligned abilities come with incorrect information about the world or contribute systematically unsound reasoning to the target, this may lead individuals and groups astray and impede their goal achievement. Furthermore, group norms (specifically those around tolerance of disagreement and dissent) may shape the impact of group socially minded intelligence on collective intelligence to the extent that they either facilitate diverse perspectives to increase collective intelligence or else encourage groupthink (Packer, 2008, 2009; Van Bavel et al., 2020). We have not included these possibilities in the definitions or equations for socially minded intelligence because research suggests that, *on average*, people who think together can be more intelligent they would be if they worked individually (Mercier & Sperber, 2018; Simoiu et al., 2019; Surowiecki, 2004; Wagner & Vinaimont, 2010). However, these factors represent potential complications to our conceptualization of socially minded intelligence that should be expanded on in future research. For example, one way to capture diverse informational and reasoning contributions might be to consider some individual-difference intelligences such as *g* as factors that *moderate* the impact of socially minded intelligence on individual and collective intelligence. Similarly, artificial agents might have an individual index of intelligence (e.g., Bringsjord & Schimanski, 2003; Mahoney, 1999) that interacts with socially minded intelligence to impact on individual and collective intelligence more broadly.

### *Improving Socially Minded Intelligence*

Building on the previous section, socially minded intelligence might be increased in several ways for both individuals and groups. A person's socially minded intelligence can be improved *without changing anything about the person* if they join a group that can provide them with social resources to achieve their goals. Similar improvements regarding individual health outcomes have been demonstrated in the 'social cure' that flows from new group memberships and increased social identification (C. Haslam et al., 2018; S. A. Haslam et al., 2018; Jetten et al., 2012). Along related lines, socially minded intelligence may also be increased by enhancing an individual's ability to build and harness social resources. This might be achieved by increasing social (Zautra et al., 2015), emotional (Kotsou et al., 2019), or cultural (Rehg et al., 2012) intelligence; or by developing relevant cognitive skills such as theory of mind (Cavallini et al., 2015) and social self-categorization (Skorich & Haslam, 2022).

Enhancing the socially minded intelligence of a given group might be achieved by improving the relationship between group members and the group. One way to do this would be to focus on group members' social identification with the group (Ellemers et al., 2004; S. A. Haslam, 2004b; van Vugt & Hart, 2004). For example, through effective identity leadership (S. A. Haslam et al., 2020; Steffens et al., 2014). Groups with a strong sense of shared social identity are likely to be more effective in dynamic situations that require action as individuals or subgroups, in particular because such groups are better able to maintain continuity across these different social structures (Slater et al., 2016; Ullrich et al., 2005), and also because shared social identity can be used to foster psychological safety (Van Bavel & Packer, 2021), meaning that such groups have greater potential to harness the power of intragroup disagreement to benefit decision-making (Fransen et al., 2020; Malone, 2018; Mercier & Sperber, 2018; Shahinpoor & Matt, 2007). To unlock this potential, it will be important to foster productive group norms (Packer, 2008, 2009; Van Bavel et al., 2020).

At the same time, it will be important for the group's socially minded intelligence to avoid structural factors that entrench a particular identity structure within the group, such as a climate of excessive individualism or collectivism (Presbitero & Langford, 2013; Triandis, 1995), or deep faultiness (overlapping differences between group members) that make subgroup identities chronically salient (Zanutto et al., 2011). If group members always see themselves as either individuals, subgroup members, or parts of a collective whole, this will reduce the extent to which they can *flexibly* see themselves and act according to a different group structure when the situation changes—because they are in a sense 'tied' to a single level of self-construal (Turner et al., 1987). Finally, collective socially minded intelligence may be improved through increasing the socially minded abilities of group members through selection or training.

## Socially Minded AI

In the previous section we discussed how the concept of socially minded intelligence can be understood, measured, and enhanced for individuals and groups. In this section we discuss how socially minded intelligence might be implemented in artificial systems in order to improve their ability to achieve their goals across different contexts.

### ***Differentiating Socially Minded Intelligence From Existing Approaches to AI***

The agent-environment framework commonly used to design artificial agents (Legg & Hutter, 2007; Russell & Norvig, 2021; see Figure 2) can be used to demonstrate how socially minded intelligence could differ from the standard conceptualization of intelligence within artificial agents. In this framework, processes such as perception and cognition occur within the individual agent. Other agents are treated as objects in the environment, unless the agent being modelled is part of a multi-agent system, in which case other agents are treated as agents in the environment (Russell & Norvig, 2021). In both cases, though, other agents are treated as existing in the environment, not as part of the agent itself. This treatment of other agents is analogous to the way that individualistic tests of human intelligence treat other people as something to be either included or excluded from the testing environment, rather than as something that *contributes to* a person's intelligence.

**Figure 2**

*The Agent-Environment Framework (Redrawn from Russell & Norvig, 2021, Figure 2.1, p. 55)*

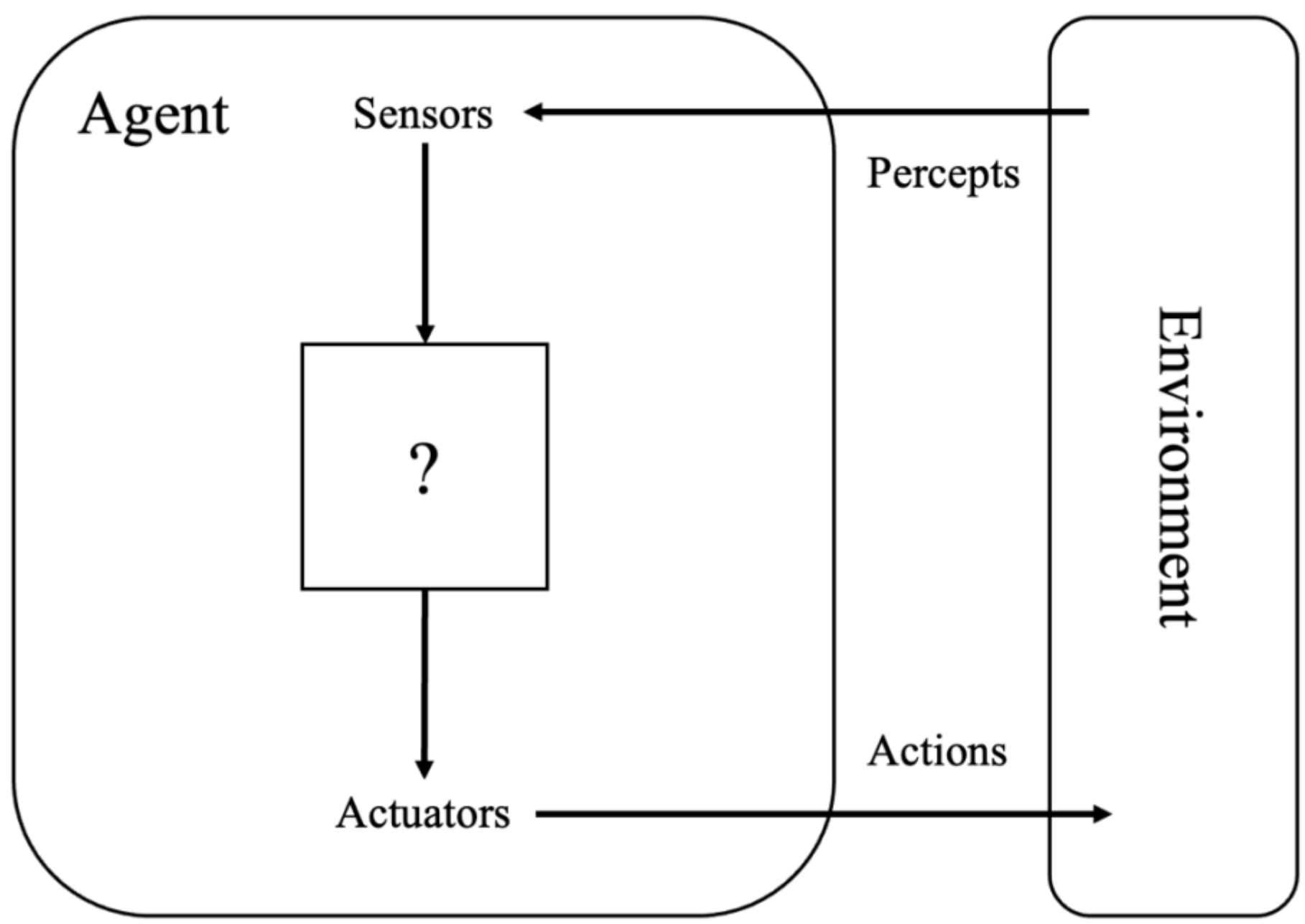

*Note.* This figure is redrawn from Russell & Norvig (2021, Figure 2.1, p. 55). It represents the standard agent-environment framework which is commonly used to design artificial agents. In such a framework, processes such as 'perception' and 'cognition' typically occur within the target agent, while other agents exist in the target agent's environment. This can be contrasted with the socially extended framework represented in Figure 3, in which other agents can form part of a socially extended version of the target agent, allowing otherwise internal processes to be shared with those other agents.

In contrast, a socially minded artificial agent goes one step further than treating other agents as 'agents in the environment', in that, where appropriate, it can include other agents *as part of itself in a broader sense*. That is, socially minded agents do not only perceive, think about, and act towards other agents; but they can also perceive, think, and act *with* other agents. In people, this can be understood through the concept of shared social identity, in which two or more people identify with the same group (e.g., Hogg & Rinella, 2018). Here there is a cognitive redefinition of the self (Turner et al., 1987) which becomes a platform for shared perception, cognition, and action. This idea can be conceptualized for artificial agents by using an extended version of the agent-environment framework, as represented schematically in Figure 3.

**Figure 3**

A Socially Extended Agent-environment Framework

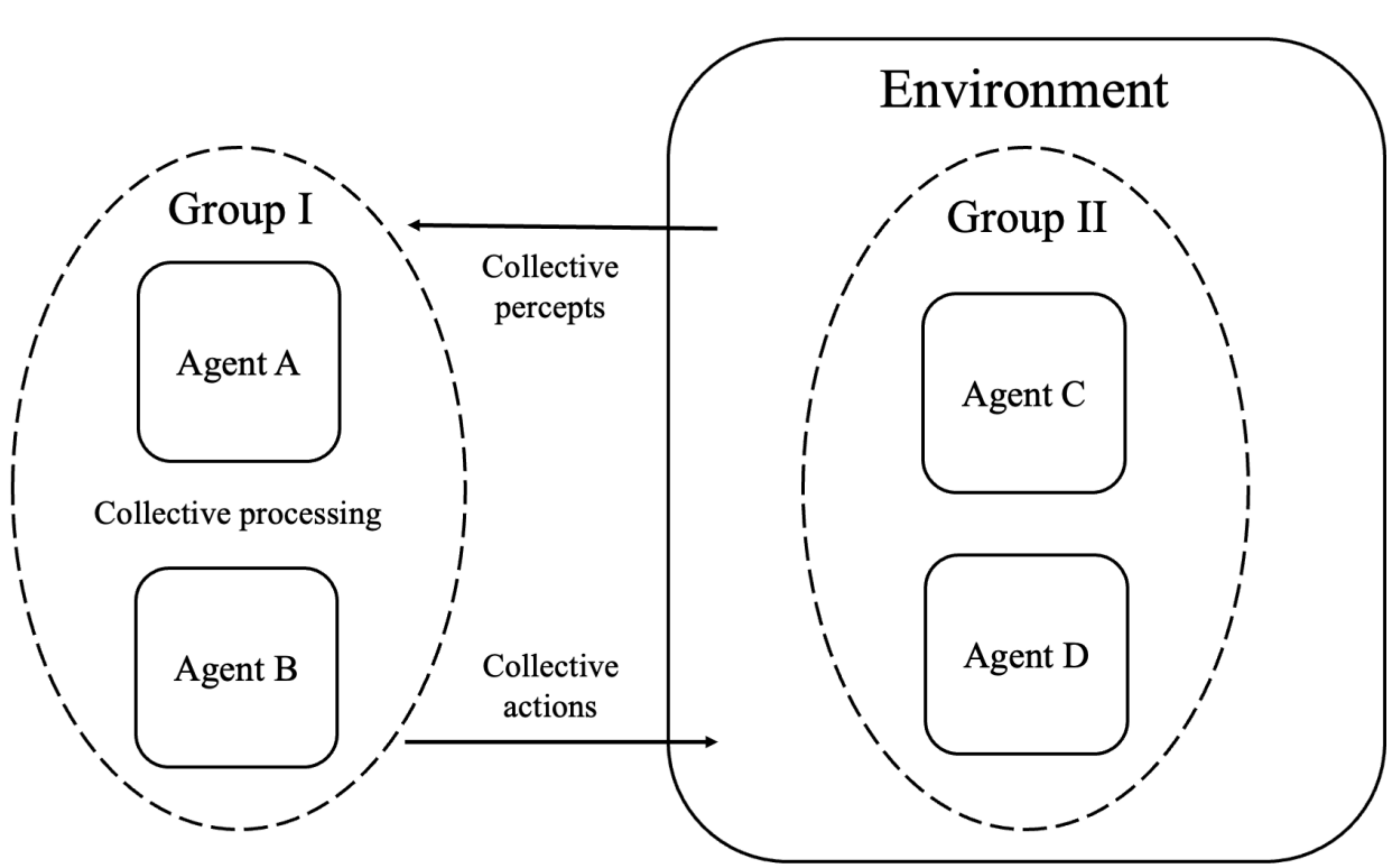

*Note.* This figure represents in a simplified manner what it means for an agent to perceive, think, and act with other agents as part of themselves (in a broader sense), not just as part of the environment. To the extent that Agent A sees itself as a part of Group I (i.e., as a group member rather than as an individual agent), it can treat information coming from Agent B as part of its own perception; share its cognition with Agent B in the sense of allowing Agent B to influence its own decision-making; and act with Agent B in the environment together as a part of Group I. However, because Agent A is not part of Group II, it will treat Agent C and D as out in the environment—that is, as things to perceive, think about, and act *on* rather than as fellow group members to perceive with, think with, and act *with*. Indeed, when Agent A sees itself as an individual rather than a group member, this is also how it will treat Agent B. Finally, when relations between Group I and Group II are competitive, Agent A will act *against* Agents C and D.

Socially minded intelligence is similar to approaches in MAS that utilize subgroups (Aydin, 2012; Dorri et al., 2018; List et al., 2008), but it differs in that traditional approaches target collective outcomes. In contrast, we are proposing that agents might also be better able to achieve their *individual* goals when they can perceive, think, and act with other agents.

The socially minded intelligence approach to agent cooperation also differs from prevalent game-theoretic approaches in that it does not require either negotiation between agents (e.g., Stankovic et al., 2012; Tang & Yi, 2023) or group-level optimality (e.g., Caillou et al., 2010; Cui et al., 2012; Rădulescu et al., 2019). Rather, socially minded agents are expected to cooperate when, intersubjectively, they see each other as part of the same higher-level 'self' (i.e., they identify each other and themselves as belonging to the same self-defining group). In Figure 3, cooperation between Agent A and Agent B results from a shared sense of self as part of Group I, which changes what 'self-interest' means for these agents (e.g., when shared group membership is reflected in the agents' utility functions). In other words, to the extent that other agents share a sense of self with the target, they have been moved from being part of the problem to be solved to being part of *problem-solving*. Critically, unlike swarm agents, socially minded agents do not *have* to act as group members to be intelligent.

Having outlined how socially minded AI differs from existing approaches, in the following section we propose ways in which artificial agents might be designed to be socially minded.

### *Building Socially Minded Artificial Intelligence*

In this section we outline capabilities required for artificial agents to be socially minded, as well as proposing means by which these capabilities might be achieved.

**Making AI Agents Socially Minded.** As suggested in Figure 3, a socially minded agent should be able to perceive, think, and act with other agents. Currently, artificial agents are unable to do this in a flexible, context-sensitive manner. However, we propose that it is possible for agents that have a basic sense of mind (performing functional processes on perceptual inputs to produce actions) and a basic sense of self (being able to pursue goals) to become more socially minded (Levin, 2022). The missing steps are for mind to be flexibly *shared* and self to be flexibly *social*. Taking inspiration from the social identity perspective on social psychology (Tajfel & Turner, 1979; Turner et al., 1987), we propose three key capabilities that, when added together and integrated into the architectures of existing agents, may allow them to be socially minded in this way:

Capability 1. *Social landscape modelling.* The target agent can model its current social landscape by identifying a parsimonious set of higher-order agentic structures (i.e., collectives of agents that can act together) that provide candidate explanations of the current social context. The model includes information about where agents are located within these higher-order structures, such that the target agent can identify which higher-order agentic structures a given agent belongs to, as well as which agents belong to each higher-order agentic structure. The questions this capability answers for the target agent are: 'What are the groups?', 'Who are the group members?', and 'Which groups do I belong to?'. This model is updated in response to changes in the social landscape.

Capability 2. *Self-relevance calculation*. The target agent can calculate the self-relevance of agentic structures (both higher-order structures that include the target agent, as well as an 'individual' structure that *only* includes the target agent) to determine which is most self-defining in the current situation. The individual self-structure represents the possibility that the target agent is defined best as an individual in this context. The question this capability answers for the target agent is whether it is best defined as an individual or as a member of a particular higher-order agentic structure in the current situation. Moreover, although the target agent might belong to multiple agentic structures, only one is treated as self-defining in a given context. In other words, while the target agent might identify through social landscape modelling that it belongs to both Group I and Group II (assuming these groups overlap), its self-relevance calculation needs to inform it whether it is best defined as a member of Group I, Group II, or as an individual agent in the present context.

Capability 3. *Social influence algorithm.* The target agent can identify which agents it should be influenced by, and incorporate those other agents into its perception, cognition, and action. Specifically, if the most self-relevant agentic structure in the present context is a higher-order collective structure:

3a) Information coming from agents belonging to the same self-defining higher-order agentic structure is treated as more relevant (i.e., more as part of the target agent's own perception of the environment) as a function of the strength of the self-relevance of that higher-order agentic structure.

3b) Other agents belonging to the same self-defining higher-order agentic structure have greater influence over the target agent's decision-making as a function of the strength of the self-relevance of that higher-order agentic structure.

3c) Actions that benefit other agents belonging to the same self-defining higher-order agentic structure—as well as actions that benefit the structure itself—are treated as having greater utility as a function of the strength of the self-relevance of that higher-order agentic structure.

However, if the target agent is best defined as an individual in the present context, the agent is not influenced by others in its perception, decision-making, or action in the manner we have described above.

Capability 1 (social landscape modelling) and Capability 2 (self-relevance calculation) allow an agent to have a sense of self as an individual, subgroup member, or superordinate group member which changes in response to their environment. Capability 3 (social influence algorithm) applies this sense of self-in-social-context to perception, cognition, and action. Together, these capabilities can be integrated with existing agent architectures that are based on the agent-environment framework to allow inputs, processes, and outputs to be shaped by elements of the social environment. Agents with these capabilities should be able to perceive, think, and act with other agents in a context-sensitive way, so as to optimize their socially minded intelligence.

While social identity processes have been modelled in artificial agents (Charness & Chen, 2020; Lobo et al., 2023), these processes have not been integrated in a way that allows an agent to be socially minded. Moreover, self-categorization theory (a cognitive extension of social identity theory; Turner et al., 1987) has not previously been used to create artificial agents, and it is this theory that provides possible ways to achieve the key capabilities we have identified. For example, the self-relevance of agentic structures to the target agent in a given context (Capability 2) could be calculated through a mechanism inspired by the principles of fit and perceiver readiness (Turner et al., 1987). Applied to human psychology, these principles are that a person will self-categorize as a member of a group rather than an individual based on the ratio of within- to between-group differences (comparative fit); the extent to which a given grouping fits observed patterns of behavior (normative fit); and the extent to which the person is predisposed to identify in terms of this group (perceiver readiness). Similar principles might be employed for an artificial agent to calculate the self-relevance of higher-order agentic structures in a given context. Self-categorization theory has also been applied to create theories of social influence and power (Turner, 1991), leadership (S. A. Haslam et al., 2020), and group dynamics (Ellemers et al., 2004) which are relevant to Capability 3. Regarding Capability 1, the principles of comparative and normative fit might be combined with factors identified by research into group perception (Lickel et al., 2000), attention (Shteynberg, 2010; Skorich et al., 2017), and agency (Rutherford & Kuhlmeier, 2013) to inspire approaches to modelling an agent's social landscape.

**Improving AI Through Social Mindedness.** Designing artificial agents to be more socially minded has the potential to make individual agents and multi-agent systems more intelligent in the sense that we have defined it in this paper. For an individual agent, being able to selectively utilize information from other agents in perception and decision-making, as well as being able to act together with other agents towards shared goals when necessary, has the potential to amplify its intelligence. Depending on how many other socially minded agents are present in the situation, this increase in potential intelligence may be manifold (in the same way that human intelligence is potentially multiplied by social cognition and action).

Designing artificial agents within a MAS to be more socially minded may also expand the applications for existing approaches. For example, different MAS structures are more or less effective in different problem-solving environments (Dorri et al., 2018). However, a group of socially minded agents need not have a pre-specified structure as they are able to act as individuals, subgroup members, or as a single collective depending on the situation. For socially minded agents, the situation *defines the structure of the group*, allowing a MAS incorporating such agents to be flexible in responding to changing environments. For example, rather than needing to operate as a single swarm, a socially minded MAS could split into subgroups or even work as individuals depending on the demands of the situation. Moreover, in contrast to the design of some swarm agents, socially minded agents need not be less individually intelligent in order to operate collectively (Altshuler, 2023; L. Rosenberg & Willcox, 2020). At the same time, in contrast to game theoretic approaches, socially minded agents can cooperate without needing to calculate equilibria. In this sense, socially minded agents may represent a middle ground between approaches to MAS which focus on coordinating 'smart but selfish' individual agents and those that make individual agents less intelligent for the sake of collective coordination.

## Socially Minded Intelligence in Other Intelligent Systems

In this paper we have defined both 'intelligence' more broadly, and socially minded intelligence specifically, in a manner that might apply to any *agent-based* intelligence. In this section we discuss the concept of socially minded intelligence in the context of other such intelligent systems.

### *Socially Minded Human-AI Teaming*

One research area that may benefit from a socially minded intelligence perspective is human-AI teaming, where humans and AI systems work together in order to be more intelligent than either could be alone (Malone, 2018, p. 241). This kind of teaming has tremendous potential but is complex and fraught, with the introduction of artificial agents potentially impairing existing team functioning (Jung et al., 2015; McNeese et al., 2018; O'Neill et al., 2022). The AI teaming challenge has been characterized as having a fundamentally *cognitive* basis, such that AI teammates need to be better able to process and respond to social information in a dynamic way (Chakraborti et al., 2017). As artificial intelligence increases, people expect AI teammates to be more socially responsive (Glikson & Woolley, 2020) and studies have found that enabling AI agents to adjust dynamically to their human teammates can improve teaming outcomes (Hauptman et al., 2023; R. Zhang et al., 2021). This kind of 'teaming intelligence' in AI agents has been conceptualized more concretely as "knowledge, skills, and strategies with respect to managing interdependence" (Johnson & Vera, 2019, p. 18), which allows a human-AI team to establish common ground, to be flexible regarding team structure, to coordinate action, and to make decisions collectively. However, currently AI agents generally lack this kind of intelligence and accordingly effective, *interdependent* human-AI teaming is rare (Hagemann et al., 2023; Johnson & Vera, 2019; Schmutz et al., 2024)

Socially minded intelligence might be applied to improve human-AI teaming. For example, the capabilities we have outlined may be useful for designing AI agents that can better respond to changes in the social situation that might lead a human teammate to act more as an individual or as a group member. Such rudimentary socially minded agents might improve human-AI teaming simply by identifying when they should work with a human teammate or work autonomously (as a function of information from their social environment). Furthermore, because the capabilities we have outlined are modelled on how social identity works in humans, to the extent that socially minded agents can effectively deploy all three of these capabilities (social landscape modelling, self-relevance calculation, and social influence algorithm), this would effectively endow them with an artificial social identity capability that would allow for a more *intersubjective* connection with humans. Such *fully-actualized* socially minded AI teammates would be more akin to 'thought partners' than tools (K. M. Collins et al., 2024). Accordingly, people may be more likely to treat AI agents with these capabilities as true teammates, which is a significant barrier to establishing team cognition in human-AI teams (Musick et al., 2021). Currently, although humans can treat robots as partners, even to the extent of favoring ingroup robots over human outgroup members (S. Collins et al., 2024; Fraune et al., 2017), they do not treat them as if they are intersubjective partners—that is, as having shared mind (E. R. Smith et al., 2021). Making AI teammates more socially minded may therefore address criticisms of static behavior and social identity based divides in current human-AI teaming (Schelble et al., 2022; R. Zhang et al., 2021).

In summary, fully actualized socially minded human-AI teams, in which both humans and AI teammates can flexibly and intersubjectively see each other as forming part of the same higher-order agentic structure (i.e., as belonging to the same group) may be more effective than existing human-AI teams—particularly in situations where it is not possible to specify roles beforehand, or where roles or team structures need to change in response to circumstances (Madni & Madni, 2018; Zhao et al., 2022). We therefore predict that increasing the capacity to be socially minded in artificial agents will not only increase the potential for human teammates to perceive that they have mind, but also increase the potential effectiveness of human-AI teams.

***Other Intelligent Systems***

We have outlined how socially minded intelligence might be conceptualized, measured, and enhanced in people, AI systems, and human-AI teams. However, there are many other kinds of intelligent systems that can achieve their individual and collective goals in changing environments, ranging from the human-like intelligence of other apes (Burkart et al., 2017; Leavens et al., 2019), to swarming and flocking social animals such as fish, birds, and insects (Bajec & Heppner, 2009; Garnier et al., 2007; Kasumyan & Pavlov, 2023), to less familiar intelligences such as fungi and single-celled slime molds (Adamatzky et al., 2023; Mayne et al., 2015; Reid & Latty, 2016). We propose that any intelligent system that can be considered an 'agent' (i.e., as perceiving its environment through sensors and acting upon that environment through actuators) can potentially have socially minded intelligence. However, we further propose that this is contingent on several factors.

First, to have individual socially minded intelligence, an agent must have some socially minded ability (traits and skills that allow it to flexibly perceive, think, and act with others). In the present paper we have outlined self-categorization ability as one way to quantify this ability, but animals that can flexibly engage in flocking or swarming behaviors by following coordination rules (Bajec & Heppner, 2009; Kasumyan & Pavlov, 2023; Sirot, 2006) may also form a kind of social mind—with the collective's response to inputs shaping the individual agent's perception, cognition and action in a distributed way (i.e., via neighboring agents). The *degree of flexibility* here determines the level of socially minded ability, with flexibility referring to both the range of possible social minds formed and the speed at which these can be formed. A key indicator of this flexibility is the extent to which individual agents can act as individuals, as subgroups, or as a superordinate group in response to changing situations. Humans have the ability not only to represent themselves hierarchically (as embedded within groups within groups; Marsh & Shavelson, 1985; Turner et al., 1987), but can also do this *recursively* (Corballis, 2007), allowing for a large number of potential social minds that can be formed in a given situation. Other apes such as chimpanzees have the ability to form subgroups and work as individuals (Badihi et al., 2022; Funkhouser et al., 2024), but they appear to have less recursive ability than humans (Corballis, 2014), and accordingly we propose that they should have less socially minded ability.

Second, the agent must have other socially minded agents to form an intersubjective social mind with, which means that a purely solitary agent (e.g., most wild *Felidae;* Bradshaw, 2016) has no socially minded intelligence. Third, those other agents must see themselves as forming the *same* social mind as the target agent (i.e., leading to self-overlap). Here, agents that are predisposed to compete rather than cooperate have a barrier to their potential socially minded intelligence. Along these lines, research suggests that, as do humans, chimpanzees use a variety of enforcement mechanisms to mitigate competition (Suchak et al., 2016). However, agents without socially minded intelligence might still cooperate for their own benefit (e.g., through reciprocal or kinship altruism; Schino & Aureli, 2010). As with both human and artificial intelligence, some degree of individual intelligence operates independently from socially minded intelligence.

Moving to collective socially minded intelligence, we propose that this additionally requires individual agents to explicitly represent or implicitly encode the group itself in some way (in line with our definition of 'group' outlined previously). Indeed, some more complex collectively intelligent systems, such as ant colonies, are made up of component agents that utilize collective representations. In particular, ants have the ability to recognize nestmates—even those from others species of ants—"via phenotype matching, where a template is compared to the label of encountered individuals" (Bos & d'Ettorre, 2012, p. 4). This template is equivalent to a representation of the group, and thus we propose that ant colonies can have a degree of group socially minded intelligence. Similarly, bee colonies have specific pheromone profiles that can be used by bees to establish shared group membership, with evidence that this is unrelated to genetic similarity (Breed et al., 2004). However, a collective that operates through simple schooling rules (e.g., 'join conspecifics of similar size'; 'imitate my neighbor'; Kasumyan & Pavlov, 2023) does *not* have group socially minded intelligence unless the school itself is represented or encoded within the individual agents. Indeed, such a collective does not constitute an agent *per se* because its members do not have a shared higher-level goal. However, this is not to say that members of such a collective may not have *individual* socially minded intelligence afforded in a distributed way via their neighbors (Theiner, 2018). For example, an individual fish might be better able to avoid a predator they cannot personally see by following a schooling rule that provides indirect information about the predator's presence. This further supports the idea that while individual and collective socially minded intelligence share a mathematical similarity (see Box 3), they are not equivalent.

An interesting case to consider is that of human–dog teams. In such teams, despite teammates being from different species, there may be a degree of shared subjectivity because dogs appear to have the ability to be socially minded (Kaminski & Marshall-Pescini, 2014) and likely co-evolved with humans in a socially integrated manner (Chambers et al., 2020), suggesting that this ability is not merely analogous but is potentially overlapping. Indeed, domesticated dogs and humans share similar, if not identical, social cognitive abilities (MacLean et al., 2017). Moreover, evidence suggests that dogs and humans can share joint intentionality (Horschler et al., 2022). Accordingly, we propose that human–dog teams may indeed have both individual and collective socially minded intelligence. This highlights the potential for developing socially minded intelligence within human-AI teams by utilizing compatible cognitive mechanisms.

**Towards Tests of the Construct of Socially Minded Intelligence**

In this paper we argue that the construct of socially minded intelligence differs from existing conceptualizations of intelligence that focus on capabilities that are either individual or collective (but never both). In particular, we propose that socially minded intelligence explains unique variance in the performance of individuals and groups, beyond existing intelligence constructs. This claim requires direct empirical evidence, and the merits of socially minded intelligence as a more complete model of intelligence could be tested in the following ways for individual people, groups, artificial agents, and multi-agent systems.

For individual people, one approach would be to identify people's personal goals and then compare performance on these goals across a variety of environments with and without social resources (i.e., other people they can work with). If people's ability to achieve their goals is affected by their social resources, socially minded intelligence explains variance above that of individual-difference measures of intelligence. Another way to test this hypothesis would be to compare goal-directed performance across social versus non-social contexts for people with higher or lower socially minded ability (as an individual-difference variable). If there is a positive relationship between socially minded intelligence ability and performance, with this being more pronounced in social contexts, this would support the idea that socially minded intelligence is a function of both the ability to harness social resources and the presence of those resources in the situation. Individual socially minded intelligence could also be directly manipulated rather than measured by making individual identity chronically salient or suppressing group identity by preventing 'switching' from one identity to another (Kawakami & Dion, 1993; Zinn et al., 2022). The third component of socially minded intelligence, goal alignment, might also be included as an additional moderator (manipulated or measured).

The construct of socially minded intelligence might be tested for human groups by identifying groups' goals, measuring group members' individual socially minded abilities as well as other factors that have been identified in the literature as relevant for group performance (e.g., S. A. Haslam, 2004a; Postmes et al., 2005; van Knippenberg & Ellemers, 2003), and then constructing scenarios where group goals are best achieved by group members acting as individuals, subgroup members, or group members. If group members' socially minded intelligence predicts improved group performance above the contribution of other individual and group-level factors, this would support the validity of the construct. Moreover, manipulating socially minded intelligence by making a particular identity chronically salient would also help to establish its validity. In particular, influencing group members to see themselves solely as individuals, subgroup members, or superordinate group members in contexts that require flexibility between these identities would be predicted to undermine socially minded intelligence, and hence group performance. As with individual socially minded intelligence, goal alignment between group members, subgroups, and the overall group may be either manipulated or measured to test whether greater alignment predicts better performance.

For artificial agents, the construct of socially minded intelligence could be tested by comparing the performance of existing agents against ones augmented with socially minded intelligence (i.e., by utilizing the capabilities we have outlined) in relation to a variety of goals in different social contexts. In addition to overall system performance, each capability can also be tested individually. That is, one could establish (a) whether agents that can integrate information from other socially minded agents into their perception as a function of shared higher-order agentic self-relevance are able to more accurately perceive their environment, (b) whether agents that are more influenced by other socially minded agents in their decision-making as a function of shared higher-order agentic self-relevance make better decisions, and (c) whether agents that include actions benefiting other socially minded agents and higher-order agentic structures in their utility functions as a function of shared higher-order agentic self-relevance are better able to act together with other agents. By varying the number of agents in these tests, it would also be possible to evaluate the prediction that performance improvements are a function of the number of other socially minded agents present in the situation. For multi-agent systems, similar tests to single agent systems could be conducted to evaluate whether socially minded agents are able to perform better on collective tasks (via improved perception, decision-making, and action) than agents which are not socially minded.

**Future Directions**

A pathway of future directions can be synthesized from the previous section and other parts of the paper. For measuring socially minded intelligence in humans, the first step for future research is to establish reliable measures for the different terms outlined in Boxes 1 and 2. In particular, it will be important to identify or create measures of socially minded ability, as well as personal and collective goals and their alignment. Because human intelligence is in part defined by goal setting and adaptation (Sternberg, 2020a), and because explicit goals do not always match implicit goals (Brunstein, 2025), this may involve some compromise of ecological validity. After this, researchers can test the ability of socially minded intelligence to explain variance in individual and group performance as outlined in the previous section. For artificial agents, the first step will be to create agents with socially minded capabilities (social landscape modelling, self-relevance calculation, and social influence algorithm). Such agents can be tested, individually and collectively, as outlined in the previous section. Once socially minded intelligence can be measured in people and implemented in agents, its influence on human-AI teaming can be investigated. In parallel, socially minded intelligence might be studied in other intelligent systems by utilizing the concept of 'aligned abilities' in Box 3 to identify whether this kind of intelligence exists in other systems and to what extent it predicts goal achievement across different environments.

A promising direction for future research is to explore whether some of the variance in human intelligence that is currently unexplained might be accounted for by better measures of individual or collective intelligence that are inspired by the concept of socially minded intelligence. For example, measures of individual intelligence that capture self-categorization ability, or measures of collective intelligence that capture shared social identity may explain variance in performance beyond that of existing approaches. We see this an intermediate step that may improve the measurement of intelligence to some degree, although we propose that a more radical reconceptualization (such as that afforded by socially minded intelligence) is required to more fully understand intelligence.

The conceptualization of socially minded intelligence in this paper represents a first approximation. Accordingly, there are several avenues for theoretical refinement. One previously mentioned is that self-overlap and goal overlap may not be independent in practice, because a person's personal goals may change when they identify with a group. This possibility is not currently captured by our equations. Another direction for future development is that the stable component of socially minded intelligence (i.e., socially minded ability) does not differentiate the ability of agents to be flexible versus inflexible in their self-definition from their ability to *choose how flexible to be* in relation to a given set of goals. It is possible that both these forms of 'flexibility' contribute independently to socially minded intelligence.

Another potential domain for future research to explore is the development of human socially minded intelligence throughout the lifespan. In particular, the social identity processes that we have identified as potentially contributing to socially minded intelligence have been shown to be present in children as young as five years old (Bennett & Sani, 2008; Nesdale & Flesser, 2001). However, there are some differences in social identity development across the lifespan, such as social identity effects being particularly pronounced for early (compared with middle and late) adolescents (Tanti et al., 2011). While some of these effects may represent changes in the social environment, more fundamental categorization abilities also have a developmental trajectory. For example, the outgroup homogeneity effect (differentiating ingroup faces more than outgroup faces) appears to develop in infants between three to nine months, while categorization based on gender and race develops between six to 12 months (M. Rhodes & Baron, 2019). Once social identity has developed, it tends to be applied more flexibly and sensitively as social complexity increases, social norms become more important, and cognitive abilities improve (Baron, 2015; Hitlin et al., 2006). This implies that socially minded intelligence might also develop along similar lines, but future research should explore this empirically.

**Constraints on Generality Statement**

It is important to note that much of the existing research into human intelligence, including papers that we have cited (e.g., Adair et al., 2013; Faraj & Xiao, 2006; Girotra et al., 2010), was conducted in the United States and Europe. The resulting limitation on the generalizability of concepts related to human intelligence has been noted by a range of intelligence researchers (Gould, 1981; Greenfield, 1997; Richardson, 2002; Sternberg & Grigorenko, 2004). Creating a more inclusive conceptualization of intelligence that can be generalized across cultures is an aim both of some intelligence researchers (Sternberg, 2020a) and of the current paper.

**Citations Statement**

Most of the research cited in this paper was written by academics working at universities in North America and Europe. Some researchers cited in this paper work in Asia (e.g., Cong & Zhou, 2023; Zhu et al., 2020) and Australasia (e.g., Legg & Hutter, 2007; Skorich & Haslam, 2022). Research from Africa, South America, and the Pacific is underrepresented both in this paper and the research fields we review. Moreover, there are demographic disparities within the different fields we have covered, with White researchers heavily represented in psychology (Buchanan et al., 2021; Roberts, 2024) and cisgender-male researchers heavily represented in computing (Bock, 2022; Menier et al., 2021).

**Conclusion**

In the present paper we have argued that, as things stand, our appreciation of the jigsaw of human and artificial intelligence is missing an important piece. This relates to the fact that approaches to intelligence which treat people either as isolated individuals or as mere component pieces of collectives fail to capture the dynamic relationships between individuals and groups that allow human intelligence to be both powerful and flexible across contexts. To address this lacuna, we introduced and fleshed out the concept of socially minded intelligence. This, we argue, has the capacity to unify existing approaches but also to open up new vistas of understanding and application.

As artificial agents become more prevalent and increasingly require interaction with humans, we would suggest that facilitating this kind of interactive intelligence in these agents is likely to be vital for efforts to ensure both the efficacy and integrity of AI systems. Informed by this framework, teams comprised of humans and socially minded AI agents have the potential to elevate their mutual intelligence still further. However, the potential of this brave new world is unlikely to be realized without sensitivity to the various issues that our analysis has raised. Indeed, it may be profoundly compromised.

**Appendix A. Table of variables, acronyms, and subscripts**

**Table 2**

*Summary of Variables, Acronyms, and Subscripts.*

| | |
|---|---|
| **Terms for an individual's socially minded intelligence (from Box 1)** | |
| ISMI | individual socially minded intelligence |
| SMA | socially minded ability |
| SR | social resources |
| SSI | shared social identity |
| GA | goal alignment |
| **Box 1 subscripts** | |
| p | target person |
| c | context |
| q | index for all others |
| **Terms for a group's socially minded intelligence (from Box 2)** | |
| GSMI | group socially minded intelligence |
| SMA | socially minded ability (same as Box 1) |
| AA | aligned abilities |
| GI | group identification |
| SO | self-overlap |
| SIGA | salient identity goal alignment |
| GO | goal overlap |
| N | number of group members |
| **Box 2 subscripts** | |
| g | target group |
| c | context (same as Box 1) |
| m | index for all group members |

**Terms for the generalized equations (from Box 3)**

In order of appearance, the terms AA, SMA, SO, GO, ISMI, SSI, GA, GSMI, GI, SIGA and N are defined as above in Boxes 1 and 2.

**Box 3 subscripts**

| | |
|---|---|
| x | target unit (x = p for an individual; x = g for a group) |
| c | context (same as Boxes 1 and 2) |
| q | index for all contributors |

## Appendix B. Worked examples for ISMI

### Worked Example B1. Calculating the contribution of two other people to *ISMI*

If a second person joins the context in Example 1, so that the target person now has SSI = (0.5, 0.3) and GA = (0.5, 0.7), then by Property P6, the second person will add to the social resources of the target person independently of the first person. We can see how this is implemented in the equation for ISMI as the *sum* of the social resources from each person. The two people together would increase the ISMI of the target person as follows (results shown to 2 significant figures):

$$
\begin{aligned}
ISMI_{pc} &= 0.7 * SR_{pc} \\
&= 0.7 * \textstyle\sum_q (SSI_{pqc} \times GA_{pqc}) \\
&= 0.7 * ((0.5 * 0.5) + (0.3 * 0.7)) \\
&= 0.7 * (0.25 + 0.21) \\
&= 0.32
\end{aligned}
$$

With the contribution of two other people, *ISMI* can vary from -2 to +2, however the full range is only seen with very high positive or negative alignments. This example shows the increase with the second person of ISMI from 0.15 (in Example 1) to 0.32.

### Worked Example B2. Calculating the contribution of five other people to *ISMI*

In a context where a target person, *p*, has an *SMA* of .7, with five other people present who have varying levels of *SSI* (.5, .3, .4, .7, .5) and *GA* (.5, .7, .3, .3, .9) with the target person, their *ISMI* would be calculated using equation (1) above as follows (results shown to 2 significant figures):

$$ISMI_{pc} = 0.7 * SR_{pc}$$

$= 0.7 * \Sigma_q (SSI_{pqc} \text{ x } GA_{pqc})$

$= 0.7 * ((0.5 * 0.5) + (0.3 * 0.7) + (0.4 * 0.3) + (0.7 * 0.3) + (0.5 * 0.9))$

$= 0.7 * (0.25 + 0.21 + 0.12 + 0.21 + 0.45)$

$= 0.87$

This example shows the steady increase of ISMI with the addition of aligned individuals.

**Worked Example B3. Calculating the contribution of five other people to *ISMI* with a low *SMA***

In a context where a target person, *p*, has a low *SMA* of .3, with five other people present who have varying levels of *SSI* (.5, .3, .4, .7, .5) and *GA* (.5, .7, .3, .3, .9) with the target person, their *ISMI* would be calculated using equation (1) above as follows (results shown to 2 significant figures):

$ISMI_{pc} = 0.3 * SR_{pc}$

$= 0.3 * \Sigma_q (SSI_{pqc} \text{ x } GA_{pqc})$

$= 0.3 * \Sigma((0.5 * 0.5) + (0.3 * 0.7) + (0.4 * 0.3) + (0.7 * 0.3) + (0.5 * 0.9))$

$= 0.3 * \Sigma(0.25 + 0.21 + 0.12 + 0.21 + 0.45)$

$= 0.37$

**Worked Example B4. Calculating the contribution of five other people to *ISMI* with low *SSI***

In a context where a target person, *p*, has an *SMA* of .7, with five other people present who have low levels of *SSI* (.1, .2, 0, .1, .1) and varying levels of *GA* (.5, .7, .3, .3, .9) with the target person, their *ISMI* would be calculated using equation (1) above as follows (results shown to 2 significant figures):

$$
\begin{aligned}
ISMI_{pc} &= 0.7 \ * SR_{pc} \\
&= 0.7 * \sum_q (SSI_{pqc} \text{ x } GA_{pqc}) \\
&= 0.7 * \sum((0.1 * 0.5) + (0.2 * 0.7) + (0 * 0.3) + (0.1 * 0.3) + (0.1 * 0.9)) \\
&= 0.7 * \sum(0.05 + 0.14 + 0 + 0.03 + 0.09) \\
&= 0.22
\end{aligned}
$$

## Appendix C. Worked examples for GSMI

GSMI in all examples is an average measure, independent of group size and varies from -1 to 1.

### Worked Example C1. Calculating a group's *GSMI* in a context with 1 group member.

Example C1 is an edge case where the group has a single individual and is included for completeness, to show that the metric is well-behaved with different group sizes. In a context where a target group, *g*, has 1 member present with an *SMA* of .7, that person's *GI* is .6, and the alignment between that person's current salient identity and the group's goals is .5, the group's *GSMI* would be calculated using Equation (3) above as follows (results shown to 2 significant figures):

$$GSMI_{gc} = (\Sigma_m (SMA_m * GI_m * SIGA_m))/N_m$$

$$= 0.7 * 0.6 * 0.5$$

$$= 0.11$$

The 'single group member' scenario presents an interesting edge case for the construct of socially minded intelligence and its operationalization. This is because while we have specified that socially minded intelligence only exists in contexts containing more than one agent, this context does have a 'second agent' in the form of the group. However, socially minded intelligence can only be calculated from the perspective of the group in this case (i.e., GSMI), because the group does not contribute by itself to the socially minded intelligence of the individual group member (i.e., ISMI) – the contribution of groups to ISMI operates through the psychology of other agents (which are absent here).

**Worked Example C2. Calculating a group's *GSMI* in a context with 5 group members.**

In a context where a target group, *g*, has 5 group members present with varying *SMA*s (.7, .8, .3, .9, .2), varying *GI*s (.6, .8, .2, .7, .5), and alignment between each group member's current salient identity and the group's goals (.5, .3, .8, .4, .2), the group's *GSMI* would be calculated using equation (3) above as follows (results shown to 2 significant figures):

$$GSMI_{gc} = (\Sigma_m (SMA_m * GI_m * SIGA_m))/N_m$$

$$= ((0.7 * 0.6 * 0.5) + (0.8 * 0.8 * 0.3) + (0.3 * 0.2 * 0.8) + (0.9 * 0.7 * 0.4) + (0.2 * 0.5 * 0.2))/5$$

$$= (0.21 + 0.19 + 0.048 + 0.25 + 0.02)/5$$

$$= 0.14$$

**Worked Example C3. Calculating a group's *GSMI* in a context with 5 group members with high SMAs.**

In a context where a target group, *g*, has 5 group members present with high *SMA*s (.8, .9, 1, .8, .9), with the same *GI*s (.6, .8, .2, .7, .5) and alignment between each group member's current salient identity and the group's goals (.5, .3, .8, .4, .2), the group's *GSMI* would be calculated using equation (3) above as follows (results shown to 2 significant figures):

$$GSMI_{gc} = (\Sigma_m (SMA_m * GI_m * SIGA_m))/N_m$$

$$= ((0.8 * 0.6 * 0.5) + (0.9 * 0.8 * 0.3) + (1 * 0.2 * 0.8) + (0.8 * 0.7 * 0.4) + (0.9 * 0.5 * 0.2))/5$$

$$= (0.24 + 0.22 + 0.16 + 0.22 + 0.09)/5$$

$$= 0.19$$

**Worked Example C4. Calculating a group's *GSMI* in a context with 5 group members and a high *SIGA*.**

In a context where a target group, *g*, has 5 group members present with varying *SMA*s (.7, .8, .3, .9, .2), the same *GI*s (.6, .8, .2, .7, .5), and a high level of alignment between each group member's current salient identity and the group's goals (.9, .7, .8, .8, .9), the group's *GSMI* would be calculated using equation (3) above as follows (results shown to 2 significant figures):

$$GSMI_{gc} = (\Sigma_m (SMA_m * GI_m * SIGA_m))/N_m$$

$$= ((0.7 * 0.6 * 0.9) + (0.8 * 0.8 * 0.7) + (0.3 * 0.2 * 0.8) + (0.9 * 0.7 * 0.8) + (0.2 * 0.5 * 0.9))/5$$

$$= (0.38 + 0.45 + 0.048 + 0.50 + 0.09)/5$$

$$= 0.29$$

**Worked Example C5. Calculating a group's *GSMI* in a context with 5 group members with high *GIs*.**

In a context where a target group, *g*, has 5 group members present with varying *SMA*s (.7, .8, .3, .9, .2), high *GI*s (.9, .9, .7, .8, .7), and varying levels of alignment between each group member's current salient identity and the group's goals (.5, .3, .8, .4, .2), the group's *GSMI* would be calculated using equation (3) above as follows (results shown to 2 significant figures):

$$GSMI_{gc} = (\Sigma_m (SMA_m * GI_m * SIGA_m))/N_m$$

$$= ((0.7 * 0.9 * 0.5) + (0.8 * 0.9 * 0.3) + (0.3 * 0.7 * 0.8) + (0.9 * 0.8 * 0.4) + (0.2 * 0.7 * 0.2))/5$$

$$= (0.32 + 0.22 + 0.17 + 0.29 + 0.028)/5$$

$= 0.20$